\documentclass[12pt]{amsart}
\usepackage{amssymb}
\usepackage{epsfig}

\title[central variational principle]{A Variational Principle for  Actions \\  
on Symmetric Symplectic Spaces}

\author[Rios \& Ozorio]{Pedro de M. Rios \ \& \ A. Ozorio de Almeida}

\address{Centro Brasileiro de Pesquisas F\'{\i}sicas \newline  
Rua Dr. Xavier Sigaud 150, Urca, 
Rio de Janeiro, RJ, 22290-180, Brasil.} 

\email{prios@math.berkeley.edu}

\def\R{{I\!\!R}}

\begin{document}

\thispagestyle{empty}

\vspace{1cm}

\begin{abstract}

We present a definition of generating functions of canonical relations, 
which are real functions on symmetric symplectic spaces, discussing some 
conditions for the presence of caustics. We show how the actions compose 
by a neat geometrical formula and are connected to the hamiltonians via a geometrically simple 
variational principle which determines the classical trajectories, discussing the 
temporal evolution of such ``extended hamiltonians'' in terms of Hamilton-Jacobi-type equations. 
Simplest spaces are treated explicitly.  

\end{abstract}

\maketitle

{\small 

\begin{center}
\end{center}
\vspace{0.5cm}

\section{Introduction}
\setcounter{page}{1}

The concepts of momentum and position set together on an equal footing  
define what is known  today as phase space \cite{21}, geometrically a differentiable manifold  $M$ 
endowed with a symplectic form $\omega$ (nondegeneracy could be relaxed  \cite{24}\cite{46}, but here we keep to the 
symplectic case). This generalizes Hamilton's formalism, as presented in
standard textbooks \cite{1}\cite{2}\cite{13}\cite{37}. Usually, however,  the phase space is  first treated as an 
euclidean $2n$-space, or even a cotangent bundle, with canonical coordinates $\{(p_i,q_i)\}$. While 
Hamilton's equation can be generically written down for each real differentiable function on any 
symplectic manifold, other important and familiar features of the canonical formalism do not always 
extend in an obvious fashion. Furthermore, although Darboux's theorem allows for a local canonical 
coordinatization of general symplectic manifolds, this is not always the most convenient one to use. 
It is therefore desirable to have a formalism which takes 
into account the specific geometry of the phase space, explicitly, as a further step of the 
``geometrical revenge'' in dynamics \cite{45}.

Actually, the importance of considering nontrivial geometries can be
seen from many different perspectives.  Although the law of
inertia sets euclidean geometry in a priviledged status, once a
system is subject to nontrivial constraints its proper phase space
geometry ceases to be trivial, generically \cite{10}\cite{14}\cite{24}. Similarly, for a
system which is invariant under a group of symmetries, 
it is often possible to eliminate redundant
degrees of freedom and the reduced phase space is also
generally nontrivial \cite{1}\cite{24}\cite{25}. Alternatively, not to mention relativistic contexts, 
one may be interested in the classical limit of quantum
systems, which are non-euclidean, as spin systems \cite{29}\cite{37}. 
Equivalently, it is convenient to have formalisms of classical dynamics 
intimately connected to some important representations used in
quantum mechanics. For instance, coherent states \cite{19}\cite{30} or, more pertinent to this 
work at hand, the ``center representation'' of operators  (cartesian Weyl
symbols) \cite{27}\cite{28}. Again, it is desirable to have these
``semiclassical'' formulations in nontrivial geometries, but
this requires previously well defined classical formalisms. 

Particularly, we need well defined generating functions for
finite canonical transformations. When $M$ is
the cotangent bundle over a configuration space $Q$, these
functions naturally take their values on $Q\times Q$. When $M$ is a 
K\"ahler manifold \cite{9}\cite{20}, whose metric and symplectic form are related
via its complex structure, one can define generating functions which are
bi-holomorphic on $M_{{\bf C}}\times \overline{M}_{{\bf C}}$, 
in which case the trajectories are complex. Despite the
utility of the complex and the configuration space actions in treating a variety of problems, 
we can greatly benefit from a formalism leading to actions which are real differentiable 
functions on nontrivial symplectic manifolds,  
in relationship to the real hamiltonian flows
obtained from a real variational principle. 

Here we present such a formalism. This work is a generalization on
concepts developed for  euclidean
spaces \cite{26}\cite{28}, or for linearized neighborhoods of general symplectic manifolds \cite{42},
dating back to the original work of Poincar\'e \cite{31}. The generating functions are real
functions on $M$ itself and not on a pair of local lagrangian coordinates. 
An argument of one such function, $m\in M$, can be viewed as the
``center'' of the canonical transformation $m_-\mapsto m_+$ which the
function generates implicitly. The corresponding variational
principle states that, for an appropriate family of paths $\nu
:[0,t]\rightarrow M$ whose endpoints are ``centered'' on $m$, the action 
\[
\left\{-\int_{\nu} h (m'(t'),t') dt' + \mathcal{ S}\!\!\!\!/_{\nu} \omega\right\} (m,t)
\]
is stationary for a classical trajectory. Here, $\{\mathcal{ S}\!\!\!\!/_{\nu}
\omega\} (m,t)$ is the symplectic area between the curve $\nu$ and
the geodesic arc, centered on $m$, returning from $\nu (t)$ to $\nu (0)$.

The restriction on the full foregoing theory is that $M$ be a symmetric symplectic space 
\cite{15}\cite{20}\cite{22}.  This means that $M$
admits of a complete affine connection such that every point $m\in M$ is the
isolated  fixed point of an
involutive symplectomorphism which coincides with geodesic
inversion at $m$.  Accordingly, the above formulation of the variational principle
is invariant at least with respect to general  transformations on $M$
preserving the affine connection and the symplectic form. 

The generalization from flat to nonflat  symmetric symplectic spaces
starts with the notion of double phase space, $DM$ \cite{28}\cite{45}. 
While in the flat case we define the reflection-translation group, from which the flat
theory develops ($\S$ 2), the equivalent construction  for 
nonflat spaces is not enough and we must use the fact that $DM$ is a symplectic
groupoid \cite{3}\cite{47}. Then, via a ``symmetric exponential map'', we view all
structures of $DM$ in (a subset of) the tangent bundle over $M$,
$TM$. There we define the notion of central groupoid ($\S$ 3) . 
The properties of the pullback symplectic form allows us to
see the graphs of canonical transformations as lagrangian
submanifolds $\Lambda$ of (subsets of) $TM$ and define, 
in $\S$ 4, (local) generating functions for finite canonical
transformations (more generally, relations), which are real functions on $M$. 
Such functions generate well defined canonical relations only when $\Lambda$ is
a graph over the zero section of $TM$. For canonical transformations,  
there is a further graphical condition. These considerations are discussed in $\S$ 5.  

The composition of two canonical transformations $\alpha_i : M \rightarrow M$ , 
as generated by such ``central actions'', is treated in $\S$ 6. Here, again we find
that despite their abstract nature, the rule for composing them is
very simple. If $f_{\alpha_i}$ is the central action for $\alpha_i$, the composed
central action for $\alpha_2 (\alpha_1)$ is 
\[
f_{\alpha_1}\!\vartriangle\! f_{\alpha_2}(m) = Stat_{(m_1,m_2)} 
\left\{f_{\alpha_1}(m_1) + f_{{\alpha}_2}(m_2) + \Delta (m,m_1,m_2)\right\}
\]
where $\Delta (m,m_1,m_2)$ is the symplectic area of the geodesic
triangle with given midpoints. On general symmetric symplectic spaces,
the importance of this function was first realized 
in the context of star product quantization \cite{50}, whose euclidean version 
has long been well established using the idea of centers and chords \cite{4}\cite{28}.
On the other hand, the above rule naturally generalizes the
result previously obtained on $I\!\!R^{2n}$ \cite{23}\cite{26}. By iterating these
compositions, in $\S$ 7, we arrive at the central variational
principle by taking  the limit of an infinite number of infinitesimal
canonical transformations  ($\S$ 8).  
Finally we discuss the temporal evolution of
such ``finite-time extensions'' of 
hamiltonians describing, in $\S$ 9, a central version of the
Hamilton-Jacobi equation and, more generally, their time
derivative with respect to any hamiltonian
flow, mixing Hamilton-Jacobi with Poisson brackets. 

While presenting these concepts we often make use of particular
spaces to illustrate the text. We have focused on the simplest two
dimensional cases: the euclidean plane $I\!\!R^2$, the torus $\mathcal{
T}^2$, the sphere $S^2$ and the noncompact hyperbolic plane $H^2$. Of
course, extending these examples to their cartesian products
$M=I\!\!R^{2n}$, $S^2\!\times\cdots\!\times S^2$,
$H^2\times \mathcal{ T}^2$, etc... is reasonably straightforward and has
not been considered here. We emphasize, however, that the
theory in principle applies to generic  symmetric  
symplectic spaces of arbitrary (even) dimensions.  
                
\section{Central Coordinates on Flat Spaces} 

Consider the euclidean plane $I\!\!R^2$, representing the very
simplest phase space of a single degree of freedom. $I\!\!R^2$ is a
group, under vector addition, and we can also identify the group
product as a free transitive action of $I\!\!R^2$ on itself, the translations: 
$T_{\vec{\xi}}(\vec{x}) = \vec{x}+\vec{\xi}$.
Further, $I\!\!R^2$ admits a natural involution  
$\mathcal{ R}_0(\vec{x})=-\vec{x}$, which represents a reflection, or rotation
by $\pi$, through the origin. Together with the identity
transformation this gives an action of $Z\!\!\!\!{Z}_2$ on
$I\!\!R^2$. We can  form the semidirect product:
$RT:=Z\!\!\!\!{Z}_2\ \ltimes I\!\!R^2$, which 
is called the reflection-translation group and  can also be seen
as a normal subgroup of the inhomogeneous symplectic group.
Then, $\mathcal{ R}_{\vec{x}}=(T_{\vec{x}}\mathcal{ R}_0T_{-\vec{x}})=
T_{2\vec{x}}\mathcal{ R}_0=\mathcal{ R}_0T_{-2\vec{x}}$ 
is the element of $RT$ which acts as reflection through the point $\vec{x}$:
$\mathcal{ R}_{\vec{x}}(\vec{x}')=2\vec{x}-\vec{x}'$.
$\{T_{\vec{\xi}}\ , \ \mathcal{ R}_{\vec{x}}\}$ satisfy:
\[
\hspace{2.3cm} T_{\vec{\xi}}T_{\vec{\xi}'}=T_{(\vec{\xi}+\vec{\xi}')} \ , \ T_{\vec{\xi}}\mathcal{ R}_{\vec{x}}=
\mathcal{ R}_{(\vec{x}+\vec{\xi}/2)} \ , \   
\mathcal{ R}_{\vec{x}}T_{\vec{\xi}}=\mathcal{ R}_{(\vec{x}-\vec{\xi}/2)} \ , \  
\mathcal{ R}_{\vec{x}}\mathcal{ R}_{\vec{x}'}=T_{2(\vec{x}-\vec{x}')}  \hspace{1.6cm} (2.1) 
\]

Now, let $(\vec{x}_-,\vec{x}_+)\in I\!\!R^2\times I\!\!R^2$ be such
that $\vec{x}_+=\mathcal{ R}_{\vec{x}}(\vec{x}_-)\Leftrightarrow 
\vec{x}_-=\mathcal{ R}_{\vec{x}}(\vec{x}_+)$. Then, $\vec{x}\in I\!\!R^2$ is called
the center of the pair $(\vec{x}_-,\vec{x}_+)$. Actually,  
$\vec{x}={\frac{1}{2}}\left(\vec{x}_-+\vec{x}_+\right)$, coinciding, for
the euclidean metric, with the midpoint of the geodesic arc joining
$\vec{x}_-$ to $\vec{x}_+$. Identifying  $\vec{\xi}=\vec{x}_+-\vec{x}_-$, 
the transformation $(\vec{x}_-,\vec{x}_+)\mapsto
(\vec{x},\vec{\xi})$ is a bijection. 

Finally, we see how the group of translations on $I\!\!R^2$ can be
interpreted as a composition of pairs: Let 
$\vec{x}_+=T_{\vec{\xi}'}(\vec{x}_{\lambda})$,
$\vec{x}_-=T_{-\vec{\xi}''}(\vec{x}_{\lambda})$. Then 
$\vec{x}_+=T_{\vec{\xi}'}T_{\vec{\xi}''}(\vec{x}_-)=
T_{(\vec{\xi} '+\vec{\xi}'')}(\vec{x}_-)$. But noticing that the action 
$\vec{x}_+=T_{\vec{\xi}}(\vec{x}_-)=\vec{x}_-+\vec{\xi}$ can be
identified by any of the sets $\{\vec{x}_-,\vec{\xi} \},\
\{\vec{x}_+,\vec{\xi} \}, \ \{\vec{x}_-,\vec{x}_+ \}$ uniquely,
we can rewrite the composition of translations as:
\[
\hspace{5.8cm} \left(\vec{x}_-,\vec{x}_{\lambda}\right)\odot \left(\vec{x}_{\lambda},
\vec{x}_+\right)=
\left(\vec{x}_-,\vec{x}_+\right) \hspace{5.2cm} (2.2)
\]

Although (2.2) was derived from the group product on $I\!\!R^2$ it can
be generalized for spaces which are not groups. In other words, we
can turn the argument around and identify (2.2) as the fundamental
algebraic structure on $M\times M$. This introduces the concept  of groupoid.

\section{Central Groupoids}
\setcounter{equation}{0}
\noindent
{\bf Definition 3.1} 
{\it Let $\Gamma ,M$ be spaces. $\Gamma$ is called a {\bf groupoid} over
$M$, denoted $\Gamma \rightrightarrows M$, if} :\\
{\bf Gd.0)} {\it $\exists$ two maps $P_-,\ P_+:\Gamma \rightarrow M$,
called the source and target maps, respectively}.\\
{\bf Gd.1)} {\it Let $(\Gamma \times \Gamma)\supset \Gamma_2 :=
\left\{(\gamma ',\gamma '')|P_+(\gamma ')=P_-(\gamma '')\right\}$.
$\Gamma_2$ is called the set of composable elements. Then, $\exists$
an associative map $``\odot $'': $\Gamma_2\rightarrow \Gamma$, 
$(\gamma ',\gamma '')\rightarrow \gamma '\odot \gamma ''$, 
called groupoid composition, or product,   satisfying
$P_-(\gamma '\odot \gamma '')=P_-{(\gamma ')}\ , \
P_+(\gamma '\odot \gamma '')= P_+(\gamma '')$ } \\
{\bf Gd.2)} {\it $\exists$ an involution $i:\Gamma \rightarrow \Gamma$, 
called inversion, s.t. $\forall \gamma \in \Gamma$, its unique inverse 
$\bar{\gamma}\equiv i(\gamma )$ satisfies:  
$\bar{\gamma}\odot (\gamma \odot \gamma ')=\gamma '$ if $(\gamma
,\gamma ')\in \Gamma_2,\ (\gamma '\odot \gamma )\odot \bar{\gamma}=
\gamma '$ if $(\gamma ',\gamma )\in \Gamma_2$ }  

{\it $M$ can be identified with the set of identities in $\Gamma$, 
$\Gamma_e:=\left\{\gamma \odot \bar{\gamma}\right.$, or 
$\left.\bar{\gamma}\odot \gamma \ | \ \gamma \in \Gamma \right\}$. Every 
$\gamma_e\in \Gamma_e$ satisfies
$P_-(\gamma_e)=P_+(\gamma_e)$, but the converse is not necessarily true}.

\vspace{0.3cm}
\noindent
{\it Examples:}
The simplest examples of groupoids are:

\noindent
{\bf\it i}) $\Gamma$ is a group  $G$, $\Gamma_e\equiv \{e \}\equiv M\ , \
\Gamma_2\equiv G\times G$

\noindent
{\bf\it ii}) The pair groupoid $\Gamma \equiv M\times M$, $P_-$ and
$P_+$ being the first and second projections. 
In this case, $\Gamma_e\simeq M$ is the diagonal, inversion is
permutation  and the compositon is 
\[
\hspace{5.5cm} (m_-,m_{\lambda})\odot (m_{\lambda},m_+)=(m_-,m_+) \hspace{4.8cm} (3.1)
\]
which generalizes (2.2). These two
examples are complementary in the sense that they have, respectively,
the minimal and the maximal identity spaces possible.

\vspace{0.3cm}
We now focus on those
groupoids $\Gamma$ which are also  symplectic manifolds and
for which groupoid composition respects its symplectic structure \cite{3}\cite{47}. 

\vspace{0.3cm}
\noindent
{\bf Definition 3.2} {\it A groupoid $\Gamma \rightrightarrows M$ is {\bf symplectic} if
$\left(\Gamma ,\omega_{\Gamma}\right)$ is a symplectic manifold and} :\\
{\bf LGd)} {\it $M\simeq \Gamma_e$ is a submanifold of $\Gamma$, $P_{\pm}$ are
submersions and $i:\Gamma \rightarrow \Gamma \ , \ \odot :
\Gamma_2\rightarrow \Gamma$ are smooth. In this case $\Gamma$ is
called a Lie Groupoid } .\\
{\bf SGd)} {\it The graph of $``\odot$'' is a lagrangian submanifold of $\Gamma
\times \Gamma \times \bar{\Gamma}$, where $\Gamma\rightarrow \bar{\Gamma}$
is an antisymplectic isomorphism: 
$\left(\bar{\Gamma} ,\omega_{\bar{\Gamma}}\right)\equiv \left({\Gamma},-\omega_{\Gamma}\right)$.
In shorthand notation, we write this as an ``additive'' property : $\omega_{\Gamma}(\gamma_1\odot \gamma_2) \ \approx \  \omega_{\Gamma}(\gamma_1)+\omega_{\Gamma}(\gamma_2)$ }. 

{\it As consequences, $M\simeq \Gamma_e$ is a lagrangian submanifold of
$\Gamma$ and inversion is anti-symplectic, i.e. the graph of $``i$''
is a lagrangian submanifold of $\Gamma \times \Gamma$ , $\omega_{\Gamma}(\bar{\gamma}) \approx  -\omega_{\Gamma}(\gamma)$ } .

\vspace{0.3cm} 
\noindent
{\it Example:} let $M$ be a sympletic manifold, $\omega$ its symplectic
structure. Then, the pair groupoid $\bar{M}\times M$, with sympletic
structure $\omega_{\Gamma}=\delta \omega$, where 
\[
\hspace{5.7cm} \delta \omega := P^{\ast}_+(\omega)-
P_-^{\ast}(\omega)\equiv \omega_+ \ominus \omega_- \ , \hspace{4.5cm} (3.2)
\]
is a
symplectic groupoid, called the {\it product} or 
{\it {\bf double phase space} } and denoted
$DM$. ~~ That is ~~ $DM \equiv (\bar{M}\times M)\rightrightarrows M$. Notice that
while $DM$ satisfies SGd, the cartesian product $M\times M$, with
the usual  
symplectic form $\omega \oplus \omega$, does not.

\vspace{0.3cm}
In order to define the notion of ``center'' precisely, we need to
consider phase spaces which are symplectic, but also
have  the property of being symmetric \cite{15}\cite{20}\cite{22}:

\vspace{0.3cm}

\noindent
{\bf Definition 3.3} {\it A differentiable manifold $M$ endowed with a
symplectic form $\omega$  shall be called a {\bf symmetric symplectic
space} if  $(M,\omega )$ admits of a complete affine connection $\nabla$ which is symplectic,
ie. $\nabla \omega = 0$,  such that 
$\forall m\in M,\ m$ is the isolated fixed point of an
involutive symplectomorphism  $\mathcal{ R}_m:M\rightarrow M$ which coincides with the
inversion, at $m$, of the geodesic flow, for all geodesics through $m$.
Accordingly, $\mathcal{ R}_m$ satisfies $\mathcal{ R}_m\mathcal{ R}_{m'}\mathcal{ R}_m = \mathcal{ R}_{\mathcal{ R}_m(m')}$ , 
$\{\mathcal{ R}_m\mathcal{ R}_{m'}\}$ is the group of displacements on $M$ \ 
and $\nabla$ is the corresponding canonical connection with null torsion and parallel curvature}. 

\vspace{0.3cm}

This generalizes the reflection-translation structure on euclidean space.  
An important particular case is when $\nabla$ {\it is the Levi-Civitta connection associated to a complete riemannian 
metric} $\eta$. In this case, $\mathcal{ R}^{\ast}_m(\eta ,\omega )=(\eta ,\omega )$ and one can show \cite{17}\cite{35} that   
$(M,\omega ,\eta )$ is a {\bf hermitian symmetric space}, that is, 
$M$ is a K\"ahler manifold \cite{9}\cite{20} whose complex structure $J$ satisfies $\nabla J=0$ and 
$\eta (JX,Y) \equiv \omega (X,Y),\ \forall X, Y \in \mathcal{ X}(M)$ , 
and for which 
the involution $\mathcal{ R}_m$ is holomorphic.
More generally, part of the following theory, namely the local definition 
of central generating functions ($\S$ 4), can be developed on 
general symplectic manifolds \cite{31}\cite{41}\cite{42}\cite{43}. 
However, the geometrical law of composition ($\S$ 6) and its 
consequences need the full setting of a symmetric symplectic space. 
As simplest examples of these, we shall consider the flat
euclidean plane $I\!\!R^2$ and torus $\mathcal{ T}^2$, both being also
groups, and as simplest nonflat examples, the sphere
$S^2$ and the noncompact hyperbolic plane $H^2$, 
both being hermitian symmetric spaces.    

When $M$ is a  symmetric symplectic space,
let us denote the exponential map by $\hspace{0.1cm}$~~~~ Exp${}_m\!:\!T_mM\rightarrow M$ , 
$\vec{v}\mapsto \rho_{\tau}(1)$ , where $\rho_{\tau}(t)$ is the geodesic in
$M$ defined by $\rho_{\tau}(0)=m$ , $\dot{\rho}_{\tau}(0)=\vec{v}$ .\linebreak
We can define a
(restricted) groupoid structure directly on (a subset of) its tangent
bundle $TM$, associated to $DM$ via the {\it {\bf
symmetric exponential map} }:
\[
\Phi \ : \ TM\rightarrow DM \quad , \quad 
\tau =(m,\vec{v})\mapsto \gamma = \left(\mbox{Exp}_m(-\vec{v})\ , \
\mbox{Exp}_m(\vec{v})\right)\ .
\]
Here, $\Phi$ is
``symmetric'' in the sense that $(\mathcal{ R}_m \circ \Phi )(m,\vec{v})=\Phi
(m,-\vec{v})$ and thus $\gamma \rightarrow \bar{\gamma}\equiv
i(\gamma )$. Denoting 
$\Omega =\Phi^{\ast}(\delta \omega )$, $\delta \omega$ defined in (3.2), we have that $\Omega$ is a closed
2-form  on $TM$ since $\delta \omega$ is a symplectic form on $DM$. If $M$ has no closed geodesics, 
as for a hermitian symmetric space of noncompact type (or $\R^{2n}$), then $\Phi$ is a bijection and
$(TM,\Omega )$ is a sympletic manifold. But otherwise, $\Phi$
is not globally invertible. Consider then the set 
$(TM)_0 \subset TM,\ (TM)_0$ being the maximal subset, connected to
the zero section $T^0M\equiv \{(m,\vec{0})\}\subset $ $TM$, on which $\Phi$
is invertible. Denote $\Phi_0 :=\Phi \big|_{(TM)_0}$, 
$\Omega_0 :=\Omega \big|_{(TM)_0}\equiv \Phi^{\ast}_0(\delta \omega )$. Then 
$\left((TM)_0,\Omega_0\right)$ is a symplectic manifold. 
Now, it is known that for complete riemannian manifolds, only the stratum of $\Sigma_1 \subset TM$ has maximal dimension, 
where $\sigma_i(m) = \Sigma_i\cap T_mM$ is the subset of $T_mM$ defined by $I(v) = i$ , $i=1,2,...$ , 
with $I(v) =$ number of geodesics of length $\| v\|$ which connect $m$ and $Exp_m(v)$ , see \cite{20.1}. 
Thus, if $M$ is a hermitian symmetric space of compact type (or the torus), 
although $\Phi_0:(TM)_0\rightarrow DM$ is not onto, it only leaves out a set
of measure zero with respect to the Liouville form on  $DM$ obtained
from $\delta \omega$. We conjecture that this may be true for other symmetric symplectic spaces 
with closed geodesics, as well.  
These facts motivate a slight modification in the concept of symplectic groupoids, suitable to our purposes: 

\vspace{0.3cm}
\noindent
{\bf Definition 3.4} {\it Let $M$ be a symmetric symplectic space.  
$\left((TM)_0,\Omega_0\right)$ shall be called the {\bf standard central
groupoid over $M$}, denoted $(TM)_0\Rrightarrow M$, satisfying } :\\
{\bf CG.O)} {\it $\exists$ three maps $P_0,P_-,P_+$ : $(TM)_0 \rightarrow M$,  called the central,
source and target maps, respectively, where  $P_0(m,\vec{v})=m$ is the natural
projection  and  
$P_{\pm}(m,\vec{v})=\mbox{Exp}_m(\pm \vec{v})$ } .\\
{\bf CG.1)} {\it On the restricted set of composable elements
$\left((TM)_0\times (TM)_0\right) \supset (TM)^2_0  := \{(\tau ',\tau '')\ \big|$ 
$P_+(\tau ') = P_-(\tau '') \ and \ (P_-(\tau '),\ P_+(\tau ''))\in Image(\Phi_0)\subset DM \}$,  
the composition $``\odot$'': $(TM)^2_0\rightarrow
(TM)_0$, satisfying (Gd.1), 
is defined by $\tau '\odot \tau ''\equiv \Phi^{-1}_0 \left(P_-(\tau '),P_+(\tau '')\right)$ }. \\
{\bf CG.2)} {\it $\exists$ an involution $i:(TM)_0\rightarrow (TM)_0$, called
inversion, satisfying all properties in (Gd.2) plus 
$P_0\left(i(\tau )\right)=P_0(\tau )$, $\forall \tau \in (TM)_0$.
Specifically, if $\tau =(m,\vec{v})$ then $\bar{\tau}\equiv i(\tau
)=(m,-\vec{v})$ }. \\
{\bf CG.3)} {\it The identiy space is the zero section and inversion is smooth. 
However, $P_{\pm}$ are only locally submersions and 
$\odot$ is locally smooth. $(TM)_0$ is a 
local Lie groupoid } .\\
{\bf CG.4)} {\it The graph of $``\odot$'' is the union of lagrangian
submanifolds of $(TM)_0\times (TM)_0\times (\overline{TM})_0$. The
graph of $``i$'' is a lagrangian submanifold of $(TM)_0\times
(TM)_0$, the zero section being  lagrangian in $(TM)_0$.
We may call $\left((TM)_0,\Omega_0\right)$ a local symplectic
groupoid } . 

\vspace{0.3cm}
When $M$ has no closed geodesics, 
$(TM)_0\equiv TM$, $\Phi_0\equiv \Phi$, $TM$ is diffeomorphic to
$DM$. In this case, the words ``restricted''  and ``local''
do not apply and the central groupoid is a bonafide
symplectic groupoid with the extra structure of a central map. 
 
\vspace{0.3cm}
\noindent
{\it Examples:} $(TI\!\!R^2)_0\equiv T I\!\!R^2$, $(TH^2)_0\equiv
TH^2$, the trivial cases since both $I\!\!R^2$ and $H^2$ have no closed 
geodesics. For  $\mathcal{ T}^2$  we have the following: Let $AS^1$
denote the set of antipodals in $S^1\times S^1$, i.e. $\{(\varphi
,\varphi \pm \pi )\}$, under the usual representation 
$S^1\ni e^{i\varphi},\ \varphi \in [0,2\pi ]$. The subset
$T\mathcal{ T}^2\supset (T\mathcal{ T}^2)_0:=\left\{\tau =(p,q\ ;\ v_p,v_q)|\ |v_p|,|v_q|
<\pi /2\right\}$  
is s.t. $\Phi_0:(T\mathcal{ T}^2)_0\rightarrow D\mathcal{ T}^2\backslash
(AS^1\times AS^1)$ is bijective. Clearly $AS^1\times AS^1$ is a set
of measure zero in $\mathcal{ T}^2\times \mathcal{ T}^2$. For 
$S^2$ the situation is very similar: Let $A S^2$ denote the set of
antipodals in $DS^2$. Taking the subset 
$TS^2\supset (TS^2)_0:= \left\{\tau =(m,\vec{v})\big |\ |\vec{v}|<\pi
/2\right\},$
then $\Phi_0:(TS^2)_0\rightarrow DS^2\backslash AS^2$ is
bijective. Again, $AS^2$ has measure zero in $DS^2$.

\vspace{0.3cm}
However, when $M$ has closed geodesics, the standard central groupoid is not the
only central groupoid possible. To see that, consider another subset
$(TM)_1 \subset TM$  which satisfies:
$(TM)_1\cap (TM)_0=\O \quad \mbox{and}\quad \Phi \left((TM)_1\right)=
\Phi \left((TM)_0\right)\equiv (DM)_0\subset DM\ .$
Furthermore, denoting the map restriction by $\Phi_1\equiv \Phi
\big|_{(TM)_1}$, we limit attention to those subsets on which
$\Phi_1$ is almost everywhere injective. In other words, the set
$N \!\subset\! (DM)_0$ on which $\Phi^{-1}_1$ is multiple valued has measure zero
with respect to the Liouville form on $DM$ obtained from $\delta \omega$. Then,
defining an equivalence relation among different pre-images in
$(TM)_1$ of the same point in $(DM)_0$, we have another bijection 
$\tilde{\Phi}_1:(\widetilde{TM})_1\rightarrow (DM)_0$, where
$(\widetilde{TM})_1\equiv (TM)_1/\sim_1$ with the
equivalence relation $\tau \sim_1\tau '$ iff $\Phi_1(\tau
)=\Phi_1(\tau ')$.
Accordingly, we denote such points in $(\widetilde{TM})_1$ by 
$\tilde{\tau} =[\tau ]_1$, where $[\tau ]_1=\left\{\tau '\in (TM)_1\
| \ \tau '
\sim_1 \tau \right\}$. If $M\subset (DM)_0$ is the 
diagonal, its pre-image $\tilde{\Phi}^{-1}_1(M) =: \widetilde{T^1M}$ is a
global cross section of $(\widetilde{TM})_1$ setting a fibration 
$P_1:(\widetilde{TM})_1\rightarrow \widetilde{T^1M}$ s.t., if
$\tilde{\tau} \in (\widetilde{TM})_1$ and $\tilde{\tau}^1 \in
\widetilde{T^1M}$ then $P_1(\tilde{\tau})=\tilde{\tau}^1$ iff 
$P_0(\tau )=P_0(\tau^1)$. In other words, the
central map $P_1$ is essentialy $P_0:TM\rightarrow M$ itself. We
denote $\Omega_1\equiv \tilde{\Phi}^{\ast}_1(\delta \omega)$. 

\vspace{0.3cm}
\noindent
{\bf Definition 3.5}
{\it We call $((\widetilde{TM})_1,\Omega_1)\stackrel{P_1}{\longrightarrow}
(\widetilde{T^1M},\omega )$ a {\bf nonstandard central groupoid}. 
On $((\widetilde{TM})_1\times (\widetilde{TM})_1) \supset (\widetilde{TM})_1^2$ 
$:= \{(\tilde{\tau},\tilde{\tau}') \big|$ 
$\tilde{P}_+(\tilde{\tau})=\tilde{P}_-(\tilde{\tau}')$ and 
$(\tilde{P}_-(\tilde{\tau}),\ \tilde{P}_+(\tau '))\in
(DM)_0 \}$, where $\tilde{P}_{\pm}(\tilde{\tau})\equiv P_{\pm}(\tau ),\ \forall \tau \in
\tilde{\tau}$,  the restricted groupoid product 
$\odot :(\widetilde{TM})^2_1\rightarrow (\widetilde{TM})_1$ is given
by $\tilde{\tau}\odot \tilde{\tau}'\equiv \tilde{\Phi}^{-1}_1(
\tilde{\Phi}_1(\tilde{\tau})\odot
\tilde{\Phi}_1(\tilde{\tau}'))$, and inversion by $i(\tilde{\tau} )\equiv
\tilde{\Phi}_1^{-1} (i(\tilde{\Phi}_1(\tilde{\tau})))$. If   
$(\widetilde{TM})_2\stackrel{P_2}{\longrightarrow} \widetilde{T^2M}$
is another central groupoid, it is
{\bf strongly equivalent} to $(\widetilde{TM})_1$ if, for every 
$\tilde{\tau}\in (\widetilde{TM})_1$ and $\tilde{\tau}' \in
(\widetilde{TM})_2$ s.t. $\tilde{\Phi}_1(\tilde{\tau})=
\tilde{\Phi}_2(\tilde{\tau} ')$, it follows $P_1(\tilde{\tau})=
P_2(\tilde{\tau}')$. If $\tilde{\Phi}_1(\tilde{\tau})=\tilde{\Phi}_2
(\tilde{\tau}')$ but $P_1(\tilde{\tau})\neq P_2(\tilde{\tau}')$,  
$\tilde{\tau}'$ is {\bf weakly equivalent} to $\tilde{\tau}$}.  

\vspace{0.3cm}
\noindent
{\it Examples:} When $M=\mathcal{ T}^2$, there are infinite 
central groupoids, but only 4 strongly inequivalent ones: 
$(T\mathcal{ T}^2)_{01} \equiv  \left\{\tau =(p,q,v_p,v_q)| \ |v_p  | <
\pi /2\ , |v_q-\pi |<\pi /2\right\}$, similarly for  
$(T\mathcal{ T}^2)_{10}$,  where  $|v_p-\pi | <
\pi /2\ , |v_q |<\pi /2$, and  
$(T\mathcal{ T}^2)_{11}$,  where $|v_p-\pi |, 
|v_q-\pi |<\pi /2 \ , $  
together with the standard one  $(T\mathcal{ T}^2)_0\equiv (T\mathcal{ T}^2)_{00}$.
In this case, all the nonstandard central groupoids are
simple subsets of $TM$, i.e. no equivalence relations had to be
considered since each $\Phi_{ij} \equiv \Phi |_{(T\mathcal{ T}^2)_{ij}}$ is
already injective. Each point in $(D\mathcal{ T}^2)_0$ has 4
strongly inequivalent pre-images: $(p,q,v_q,v_q),\ (p\pm \pi\ , q, \
v_p\pm \pi ,\ v_q)\ , \ (p,q\pm \pi ,\ v_p,v_q\pm \pi)$ and 
$(p\pm \pi \ , \ q\pm \pi \ ,\ v_p\pm \pi \ , \ v_q\pm \pi )$.  In
the case of $S^2$, there are only 2 strongly
inequivalent central groupoids. The standard one and a nonstandard
central groupoid:
$(\widetilde{TS^2})_1\equiv (TS^2)_1/\sim_1\ $, where 
$(TS^2)_1\equiv \left\{(m,\vec{v})\big| |\vec{v}|\in (\pi /2\ , \pi
]\right\}$. 
The equivalence relation is nontrivial only when $|\vec{v}|=\pi$,
coinciding with the definition of the nonstandard pre-image of the
diagonal,  
$\widetilde{T^1S^2} := \{ \ [\tau^1]_1$ ,  \ where \ 
$\tau^1=(m,\vec{v}_1)\in (TS^2)_1$ \ is s.t \ $|\vec{v}_1|=\pi$ \
 and \ $(m,\vec{v}_1)\sim_1 (m',\vec{v}'_1)$ \    
iff \ $m=m' \ \}$.   
Thus, each point on $(DS^2)_0$ has two strongly inequivalent
pre-images. If we denote $\Phi^{-1}_0(m_-,m_+)=(m,\vec{v})\equiv (\theta
,\varphi ;v,\beta )$, see below for definition of the
fiber coordinates $(v,\beta )$, then $\tilde{\Phi}^{-1}_1(m_-,m_+)=
(\pi -\theta ,\varphi -\pi ;\pi -v ,\beta -\pi)$. 

\vspace{0.3cm}
The reason for considering such nonstandard pre-images in $TM$ is, of
course, that when $M$ has closed geodesics, the geodesic arc connecting two points
in $M$ is not unique. Accordingly, the standard central groupoid
refers to the shortest of such geodesic arcs and the $k$-nonstandard
central groupoid refers to the $k$-fold ones.  Here, a $k$-fold geodesic from $m'$ to $m''$ 
is defined as the geodesic
$(m'\rightarrow m)\ast L_k(m)\ast (m\rightarrow m'')$, where $L_k(m)$ is a $k$-fold geodesic
loop based at $m$, the midpoint of the short geodesic, and $\ast$ denotes free product. 
(If $k\equiv 0$, $L_0\equiv id$ is the trivial
loop and $(m'\rightarrow m'')$ is the short geodesic). For instances of
nontrivial geodesic loops, on $\mathcal{ T}^2$, $L_k\equiv
L_{(a,b)}\equiv L^a_p\ast
L^b_q$, where $L_p$, $L_q$ are the single irreducible circuits, while
on $S^2$, $L_k$ is a $|k|$-repetition of a geodesic meridian and
it is easy to see that in this case we only need to consider strongly inequivalent
geodesics. We also consider : 

\vspace{0.3cm}
\noindent
{\bf Definition 3.6} {\it Let  $(\widetilde{TM})^2_{ij} :=   
\{(\tilde{\tau},\tilde{\tau}')\in (\widetilde{TM})_i
\times (\widetilde{TM})_j \ \big| \ \tilde{P}_+(\tilde{\tau})=
\tilde{P}_-(\tilde{\tau}')$ and $(\tilde{P}_-(\tilde{\tau}),
\tilde{P}_+(\tilde{\tau}'))$ \linebreak  $\in (DM)_0\}$. We  define
 generalized, or {\bf mixed compositions} (which are still restricted
in the original sense) as
$c^k_{ij}:(\widetilde{TM})^2_{ij}\rightarrow (\widetilde{TM})_k$~ ,~ by 
$c^k_{ij}(\tilde{\tau},\tilde{\tau}')\equiv \tilde{\Phi}^{-1}_k
(\tilde{\Phi}_i (\tilde{\tau}) \odot \tilde{\Phi}_j (\tilde{\tau}'))$.   
If $M$ is not simply connected, we may want to consider only
mixed compositions whose geodesic triangles of
composition are reducible circuits, in which case the triple
$(i,j;k)$ is called {\bf reducible}. When $M$ is
simply connected, it is enough to consider mixed compositions
within the set of strongly inequivalent central groupoids}. 

\vspace{0.3cm}
Now, the application of central groupoids in classical
dynamics rests on the following : 

\vspace{0.3cm}
\noindent
{\bf Definition 3.7} {\it Let $N$ be a symplectic space. A {\bf polarization}
on $N$ is an intergrable foliation of $N$ by lagrangian leaves } .

\noindent
{\bf Lemma 3.1} {\it $P_0$ sets a polarization on
$\left((TM)_0,\Omega_0\right)$, called the vertical or central polarization} .

\vspace{0.1cm} 
\noindent
{\it Proof:} First, every fibration is an integrable
foliation. Second, since each fiber is (an open subset of) the
tangent space at each point in $M$, the fiber dimension is half the  total
dimension of $(TM)_0$. To prove isotropy with
respect to $\Omega_0=\Phi^{\ast}_0(\delta \omega)$,
note that inversion is anti-symplectic in $(TM)_0$, i.e. 
$i^{\ast}(\Omega_0)=-\Omega_0$. But, denoting the 
fiber restriction by $(T_mM)_0$, we have that  
$i_m:=i\big|_{(T_mM)_0}\equiv \Phi^{-1}_0\circ \mathcal{ R}_m\circ \Phi_0$.
Since $\omega$ is $\mathcal{ R}_m$ invariant, i.e. $\mathcal{
R}^{\ast}_m(\omega ) = \omega$, it follows that $i^{\ast}_m\left((\Omega_0)_m\right)=
(\Omega_0)_m$, where $(\Omega_0)_m\equiv \Omega_0\big|_{(T_mM)_0}$.
Thus, $(\Omega_0)_m\equiv 0$ . \ $\square$  

\vspace{0.3cm}
{\it Further : we realize that Lemma 3.1 applies to every
$P_i:(\widetilde{TM})_i\rightarrow \widetilde{T^iM}$, as well} . 

\vspace{0.3cm}
Thus, in what follows we should consider all possible central groupoids
similarly. However, we shall mostly deal with the 
standard  central groupoid, both to simplify the treatment and
because we'll often focus on the limit $|\vec{v}|\rightarrow 0$,
which can only take place in $(TM)_0$. Accordingly, we  
often drop the denomination ``standard'' in  following definitions and discussions.  

Finally, we should compare the central groupoids with the local isomorphism 
$DM \rightarrow T^{\ast}M$, in a neighborhood of the diagonal in $DM$, 
for any symplectic manifold $M$ \cite{41}\cite{42}\cite{43}. The differences lie in the scope and 
properties of the symplectic structure. While $((\widetilde{TM})_i,\Omega_i)$ contains explicit information 
on the affine geometry (geodesic structure) of $M$, the canonical symplectic form on $T^{\ast}M$ does not.

\vspace{0.3cm}
\noindent
{\it Formulae:}
To base the abstract concepts 
of this paper, we'll often illustrate them in the simplest spaces  
$M=I\!\!R^2, \mathcal{ T}^2, S^2, H^2$. We now
provide some useful  local formulae for their 
standard central groupoids. Start with the flat spaces, locally
identical. Take  coordinates on $I\!\!R^2$ as canonical 
pairs, with usual metric
and symplectic form.
The tangent bundle is also a linear space :
$TI\!\!R^2 \owns \vec{\tau} = (\vec{x}\ ; \ \vec{v}) =
(p,q\ ; \ v_p,v_q) \ $, and 
the symmetric exponential map is :
\[
\hspace{4.7cm} \Phi (\vec{\tau}) \equiv \Phi (\vec{x}\ ; \ \vec{v}) =
(\vec{x}-\vec{v}\ ;\ \vec{x}+\vec{v}) \equiv (\vec{x}_-,\vec{x}_+) \hspace{3.8cm} (3.3)
\]
with pull-back symplectic form given by
\[
\hspace{5.8cm} \Omega = 2d\Sigma \quad , \quad \Sigma = v_pdq-v_qdp \ .  \hspace{5cm} (3.4)
\]
For the sphere, take local polar coordinates:
$S^2 \owns m = (\theta ,\varphi )\ ,  \theta \in [0,\pi ]\ ,
 \varphi \in [0,2\pi ]$~~.
We shall be using the following abbreviations: 
\[
S_{\alpha}\equiv sin
(\alpha ) \ , \ C_{\alpha} \equiv cos (\alpha ) \ , \ T_{\alpha}
\equiv tan (\alpha ), \ \ \mbox{with} \ \ S^{-1}(f), C^{-1}(f), T^{-1}(f)
\]
denoting their respective inverses. With usual metric and
symplectic form, the natural coordinates on the tangent bundle are
$TS^2\owns \tau = (\theta ,\varphi \ ;\ \dot{\theta},\dot{\varphi})$,
but it is more convenient to introduce polar coordinates on the
fibers as well, $\tau = (\theta ,\varphi \ ;\  v,\beta )$ where:
$v = |\vec{v}| = \sqrt{\dot{\theta}^2+S_{\theta}^2\dot{\varphi}^2} \ ,
\ \
vC_{\beta} = \dot{\theta} \ , \ vS_{\beta} = S_{\theta}\dot{\varphi}$
 . \ 
With these local coordinates we  write the symmetric
exponential map as
\[
\hspace{2cm} \left.
\begin{array}{l}
\Phi_0 (\theta ,\varphi ; v,\beta ) =
(\theta_-,\varphi_-;\theta_+,\varphi_+) \ \ ,\ \
\theta_{\pm} = C^{-1}(C_{\theta}C_v \mp S_{\theta}S_vC_{\beta})\ ,  \\
\varphi_{\pm} = \varphi \pm T^{-1} (S_vS_{\beta}/g_{\pm})+(1-Sign
(g_{\pm}))\pi /2 \ , \ 
\ g_{\pm} = S_{\theta}C_v
\pm C_{\theta}S_vC_{\beta}
\end{array}  
\right\} \hspace{1.5cm} (3.5)
\]
and the pull-back symplectic form $\Omega_0\equiv \Phi_0^{\ast}(\delta \omega )$
is given by
\[
\hspace{5.2cm} \Omega_0 = 2d\Sigma \ , \ \Sigma = S_{v}
(C_{\beta}S_{\theta}d\varphi - S_{\beta}d\theta ) \ . \hspace{4.6cm} (3.6)
\]
On $H^2$, we  adapt the local spherical formulas by letting
\[
\hspace{6.6cm} \theta \longmapsto i\rho  \ \ , \ \ v \longmapsto i\mu  \hspace{6.1cm} (3.7)
\]
where $i=\sqrt{-1}$. Thus,
$S_{\theta} \longmapsto \ i\tilde{S}_{\rho} , \ C_{\theta} \longmapsto
\ \tilde{C}_{\rho} \ , \ S_v\longmapsto i \tilde{S}_{\mu} , \ C_v \longmapsto
\tilde{C}_{\mu}$ , with
\[
\tilde{S}_{\alpha}\equiv sinh (\alpha ) \ , \ \tilde{C}_{\alpha}
\equiv cosh (\alpha ) \ ,  
\tilde{T}_{\alpha} \equiv tanh (\alpha
),\ \ \  \mbox{and} \ \ \  \tilde{S}^{-1}(f), \ \tilde{C}^{-1}(f), \ \tilde{T}^{-1}(f)
\]
the respective inverses.
Then, from (3.5) and (3.6) we get $\Phi$ and $-\Omega$ for $H^2$.

We can see explicitly from (3.4), (3.6) and (3.7) that the vertical
spaces are isotropic in these examples. The same holding for the zero section
$T^0M\simeq M$. Notice also  that the
pull-back symplectic form is the exact derivative of a symplectic
potential without any vertical differential components. This fact
shall be thoroughly exploited in what follows. 

\section{The Central Equation}
\setcounter{equation}{0}

We started by emphasizing the algebraic structure on double phase
spaces, or on central groupoids. Historically, the symplectic
structure was predominant, however, for it introduced the very useful concept
of action, or generating function  
of a canonical transformation on the original, or simple phase space
$(M,\omega )$. Let $\alpha$ be such a symplectomorphism $M\rightarrow M$, $\alpha^{\ast}(w)=w$.
Its graph $\mathcal{ L}_{\alpha}$ in the double phase space,  
$DM \supset \mathcal{ L}_{\alpha}: = \{ (m_-,m_+)\ |\ m_+ =\alpha (m_-)\}$, is
a lagrangian submanifold for the symplectic form $\delta \omega$, i.e.
$\delta \omega |  \mathcal{ L}_{\alpha}\equiv 0$. Similarly, if $\mathcal{
L}_{\alpha} \subset \ \mbox{Image} (\Phi_0)\equiv (DM)_0 \subset DM$,
then its pre-image $\Lambda_{\alpha}:= \Phi_0^{-1} (\mathcal{
L}_{\alpha})$ is lagrangian in the central groupoid $((TM)_0,
\Omega_0)$. Conversely, every lagrangian submanifold $\Lambda_{\alpha}:= \Phi_0^{-1} (\mathcal{
L}_{\alpha})$ defines a  {\it symplectic} or {\it {\bf canonical relation}} on $M$ \cite{43}\cite{45}, which is 
a canonical transformation when $\mathcal{ L}_{\alpha}$ is a graph over
$M_-\subset DM$. Generically, we consider those subsets $\mathcal{
L}_{\alpha}^{(r)} \subset \mathcal{ L}_{\alpha}$, satisfying $\mathcal{
L}_{\alpha}^{(r)} \subset \ \mbox{Image} (\Phi_0)$, and their
corresponding pre-images $\Lambda_{\alpha}^{(r)}$ in $(TM)_0$. 

Now, the definition of local generating functions for lagrangian submanifolds 
depends on the choice of a local
{\it symplectic potential} (a local 1-form whose derivative is
the symplectic form) suitable to a polarization which is, at least
locally, a fibration over  a referential lagrangian submanifold
containing the supports of those generating functions \cite{38}\cite{41}\cite{42}\cite{43}\cite{45}. 
For the central groupoid over $M$, we take $M$ itself,
also seen as the zero section, as the
referential lagrangian submanifold corresponding to the central
polarization.  The suitable symplectic potential is given by :

\vspace{0.3cm}
\noindent
{\bf Definition 4.1} {\it Let $((TM)_0,\Omega_0)$ be the standard central
groupoid over $(M,\omega )$. A symplectic potential $Z_0$,  for
 $\Omega_0$, shall be called a standard  
{\bf central potential} if it satisfies}
\[
\hspace{4.7cm} X \rfloor Z_0 = P_{0}' (X) \rfloor Z_0 \quad \ , \ \forall X \in \mathcal{ X}((TM)_0) \ .   \hspace{4.3cm} (4.1) 
\]
{\it Here,
$P_0'$ is the differential of $P_0$ and $\vec{x}\rfloor \alpha$ 
denotes the vector-form contraction}.

\vspace{0.3cm}
Of course, any $Z=Z_0+dQ$, $Q\in \mathcal{ C}^k_{I\!\!R} ((TM)_0)$, is another
potential, but not generally central. Condition (4.1) 
tells us that central potentials have no vertical
differential components, so we can identify these potentials
explicitly in our examples as $Z_0 = 2\Sigma$, from equations (3.4),
(3.6) and (3.7). 

In these particular examples, $Z_0$ is a global
potential on $(TM)_0$ i.e., $\Omega_0\equiv dZ_0$ is exact.  We argue that the
general case follow these known examples.
First, notice that such a
potential always exists in a small neighborhood of the zero section
$T^0M$. For, take the exact 2-form $\dot{\omega}$ on $TM$ defined by
$\dot{\omega}(m,\vec{v}):= d(\vec{v} \rfloor \omega )$, $\forall \vec{v}\in \mathcal{ X}(M)$. 
Then, $\dot{\omega}\equiv L_{\vec{v}}(w)$, the
Lie derivative of $\omega$, since $\omega$ is closed. On the other hand, when
$|\vec{v}|\equiv t \stackrel{\sim}{\rightarrow} 0$ ,  $\Omega_0
(m,\vec{v}) \stackrel{\sim}{\rightarrow}  2\lim_{\atop {t\rightarrow 0}\{\frac{1}{t}}\int^t_0  
L_{\vec{v}} (\omega ) dt' \}\simeq 2 L_{\vec{v}}(\omega )$.  That is $\Omega_0 \stackrel{\sim}{\rightarrow} 
2\dot{\omega}=2d(\vec{v} \rfloor \omega )$. Thus, in a neighborhood of $T^0M$,
$\Omega_0 \stackrel{\sim}{\rightarrow}  2d\dot{\zeta}$, where
$\dot{\zeta}(m,\vec{v}):=\vec{v} \rfloor \omega (m)$ clearly satisfies (4.1).   
(on the flat examples, \  $Z_0\equiv 2\dot{\zeta}$~~~ globally, 
but generally $2\dot{\zeta}$ only approximates the central
potential in a small neighborhood of $T^0M$~~ ). Finally,
$T^0M$ is lagrangian for $\Omega_0$ and  each vertical fiber is also lagrangian, as well as 
contractible. Thus, $\Omega_0\equiv dZ_0$ is exact.  

Recalling that
$\frac{1}{2}Z_0
(m,\vec{v})\stackrel{\sim}{\rightarrow} \dot{\zeta} (m,\vec{v})\equiv
\vec{v}\rfloor \omega$,   as $|\vec{v}|\rightarrow 0$, we may refer to the 1-form
$\dot{\zeta}$ as {\it Hamilton's potential}, since it is intimately
connected to {\it Hamilton's equation.} To see this, remember
that we can write the latter as a map \ $\dot{z}[h]:M\rightarrow TM$,
$\forall h\in \mathcal{ C}^k_{I\!\!R} (M)$, by $m \mapsto \tau_h$, where
$\tau_h=(m,\vec{v}_h)$ and $\vec{v}_h$ is given by
$(dh+\vec{v}_h \rfloor \omega )(m)=0$, or equivalently, $\dot{\zeta} (\tau_h)=-dh(m)$,
$\forall m \in M$. Or still, by denoting
$TM\supset \dot{\Lambda}_h:=\mbox{graph}$ of $\dot{z}[h]$, hamilton's
equation becomes $\dot{\zeta}|_{\dot{\Lambda}h}=-dh$, implicitly
defining $\dot{z}[h]$. 

Similarly, {\it the existence of a central potential for $\Omega_0$ allows
for a ``finite time'' extension of Hamilton's formalism }, now in the
context of generating functions. Thus, suppose that a lagrangian
submanifold $\Lambda_{\alpha}\subset (TM)_0$ is locally a graph over
$T^0M\simeq M$, that is: 
\[
\hspace{2.2cm} \mbox{Rank} (P_{0}'\ | \ T_{\tau}\Lambda_{\alpha}) = 2n = dim (M) \ ,
\
\forall \tau \in \Lambda_{\alpha} \ \ s.t. \ \ P_0(\tau ) = m\in U\subset M  \hspace{1.8cm} (4.2)
\]
then, since $dZ_0|\Lambda_{\alpha}=0$, from (4.1) we obtain: 

\vspace{0.3cm}
\noindent
{\bf Proposition 4.1} {\it For every lagrangian submanifold
$\Lambda_{\alpha}\subset (TM)_0$ satisfying (4.2) 
there exists a standard  
central generating function $f_{\alpha}\in \mathcal{ C}^k_{I\!\!R} (M)$,
satisfying the standard central equation} 
\[
\hspace{5cm} Z_0\Big|_{\Lambda_{\alpha}} =\  df_{\alpha} \quad , \quad \mbox{on}
\quad (TU)_0\subset (TM)_0 \ .  \hspace{4.2cm} (4.3)
\]
{\it Conversely, for a given $f_{\alpha}$, the above equation defines
$\Lambda_{\alpha}$ implicitly, i.e. it provides a standard section 
$z[f_{\alpha}]\equiv F_{\alpha}:U\rightarrow (TU)_0$ which is well
defined when condition (4.2) is satisfied  and as long as $f_{\alpha}$ satisfies 
appropriate standard consistency conditions } .

\vspace{0.3cm}
We shall see examples of such consistency conditions shortly. Now, 
the map $F_{\alpha}$ can be multiple
valued if $(T_mM)_0\cap \Lambda_{\alpha}$ is not unique, in which
case we should break $\Lambda_{\alpha}$ into branches
$\Lambda_{\alpha}^{(r)}$, each one uniquely given by a map
$F_{\alpha}^{(r)}$ in a subset of $M$ via
$Z_0|_{\Lambda_{\alpha}^{(r)}} = df_{\alpha}^{(r)}$, for each
$f_{\alpha}^{(r)}$, except for $\{\tau_k\}\subset \Lambda_{\alpha}$
where Rank $(P_{0}'\ | \ T_{\tau_k}\Lambda_{\alpha})< 2n$. Then, {\it via the 
symmetric exponential map, 
(4.3) locally generates a lagrangian submanifold $\mathcal{
L}_{\alpha} \subset DM$, a canonical relation on $M$}. 

{\it The similarity between Hamilton's equation and the central equation is
striking }. However, the former provides
{\it infinitesimal} transformations while the latter generates
{\it finite} relations. For this reason, 
 not every canonical relation can be generated by a real
function on $M$ via the central equation, everywhere. Generically, the
presence of {\it central catastrophes}, $\{\tau_k\}\subset
\Lambda_{\alpha}$ s.t. Rank
$(P_{0}' \ |\ T_{\tau_k}\Lambda_{\alpha})< 2n$, is unavoidable.
Their projections, $\{m_k\}\equiv \{P_0(\tau_k)\}\subset M$, are
called {\it {\bf central caustics}}. 
To circumvent this problem, new sets of ``complementary'' generating
functions are needed \cite{26}. Alternatively, we can let the 
functions depend on extra parameters and look for their 
stationary points \cite{7}\cite{8}\cite{40}. 

On the other hand, by introducing a real
parameter $\lambda = t/2$ (a scale) in the map $\dot{z}[h]$, for instance, by
multiplying every hamiltonian $h$ by $t/2$, we can see Hamilton's
equation as a map from $M$ into a small neighborhood of the
zero section in $TM$, if $t$ is sufficiently small. Since in such a
neighborhood    $\frac{1}{2}Z_0$ and $\dot{\zeta}$ are approximately equal,
$f_{\alpha}=-th$ is a central generating function for the infinitesimal
canonical transformation generated by $h$ via Hamilton's equation.
Since these are always well defined, every
infinitesimal canonical transformation can be generated by a central
function. This is obvious if we notice that such
transformations are small deformations of the identity, 
associated to lagrangian submanifolds in $(TM)_0$ which
are small deformations of the zero section, thus
satisfying (4.2).

Furthermore, the map $F_{\alpha} = -t H:M\rightarrow (TM)_0$~~, obtained
via the central equation from function $f_{\alpha}=-th$, where $h$ is
the hamiltonian, is given by $m\mapsto (m, \frac{\varepsilon}{2}
\vec{v}_h (m))$, for sufficiently short times $t\simeq \varepsilon$.
Since $\frac{\varepsilon}{2} \vec{v}_h\rightarrow 0$~~, as
$\varepsilon \rightarrow 0$~~, for very short times $t\simeq
\varepsilon \rightarrow 0$, the implicit canonical transformation
$(m_-,m_+)$ obtained via the symmetric exponential map coincides with
the linearized version $(\vec{x}-\frac{\varepsilon}{2} \vec{v}_h,\vec{x} + 
\frac{\varepsilon}{2} \vec{v}_h )$, for any choice of local linear
coordinates on $U\subset M$, s.t. $m\simeq \vec{x}$, ~~ regardless of the
specific affine geometry of $M$. Thus, {\it for very short time motion, the
geodesic segment centered on $m$ converges onto the hamiltonian orbit
that propagates from $m$ forwards and backwards in time}. 

Now, the pertinent remark should be made that the concept of generating
function is traditionally defined on the double phase space $DM$,
but this usually requires
us to previously take polarizations on $M$  itself. For general
cotangent bundles, with vertical polarization, 
the generating functions naturally take their values from pairs of
base space points, $f\equiv f(q_-,q_+)$. For general
K\"ahler manifolds, with complex polarizations,
the natural generating functions are bi-holomorphic functions
$f\equiv f(z_- , z_+)$, $z_{\pm}$ complex. {\it The present approach allows  us
to consider real generating functions on a general symmetric
symplectic space  $M$ itself}, defined via a
real polarization on the central groupoid. {\it These functions  can sometimes be pictured as ``finite time
hamiltonians'' }. More generally, {\it such central 
generating functions can locally be defined on any symplectic manifold } by 
considering a local polarization of $DM$ which is transversal to the diagonal 
$M \subset DM$. If linear coordinates are chosen on a neighborhood of a point in $M$,
these functions can be mapped to Poincar\'e's generating functions \cite{31}\cite{42}.
{\it However, the full geometrical properties of the central generating functions  
need the symmetric symplectic setting}. 

Also, we must point out that other versions of the central
equation are available :
 
\vspace{0.3cm}
\noindent
{\bf Proposition 4.1'} {\it When $M$ has closed geodesics, we can choose nonstandard central potentials  $Z_i$ s.t.} 
\[
\hspace{3.2cm} dZ_i=\tilde{\Phi}^{\ast}_i (\delta \omega ) \  \ \mbox{and} \ \ 
X \rfloor Z_i = P_{i}' (X) \rfloor Z_i  \ \ , \ \
\forall X \in \mathcal{ X}( (\widetilde{TM})_i ) \ . 
\hspace{2.8cm} (4.1')
\]
{\it Each lagrangian submanifold $\Lambda^i_{\alpha}=\tilde{\Phi}^{-1}_i (\mathcal{ L}_{\alpha}) \subset (\widetilde{TM})_i$,  
satisfying}
\[
\hspace{4.4cm} Rank(P_i'|T_{\tilde{\tau}}\Lambda^i_{\alpha})
 = 2n \ , \ over \
U\subset M \simeq \widetilde{T^iM} \ , \hspace{4cm} (4.2') 
\]
{\it is generated by a nonstandard central function $f_{\alpha}^i$ via a nonstandard central equation} :
\[
\hspace{5.4cm}Z_i|_{\Lambda^i_{\alpha}} = df^i_{\alpha} \ , \ \mbox{on} \
(\widetilde{TU})_i \subset (\widetilde{TM})_i \ , \hspace{4.9cm} (4.3')
\]
{\it which provides a nonstandard section $F^i_{\alpha}:U\rightarrow
(\widetilde{TU})_i$ defining  $\Lambda^i_{\alpha}$. When no distinction is made, 
or by means of generalization, 
we  refer to (4.3) or (4.3') simply as the {\bf central equation}} .

\vspace{0.3cm}
Notice that we have abbreviated the standard notation:
$f^0_{\alpha}\equiv f_{\alpha}$, $\Lambda^0_{\alpha}\equiv
\Lambda_{\alpha}$, $F^0_{\alpha}\equiv F_{\alpha}$, in previous
definitions. Also, the index ``$i$'' in (4.3') is not the
same as the index $(r)$ defined earlier, referring to different
branches of a single pre-image of the set $\{\mathcal{
L}_{\alpha}^{(r)}\}\subset \mathcal{ L}_{\alpha}\cap (DM)_0$. Thus,  
a generic central generating function can carry up to
two indices $\{i,(r)\}$, in order to be fully identified. 

Finally, since central potentials are defined modulo exact differentials on $M$, it is important to emphasize that {\it each nonstandard central equation}, with its appropriate nonstandard consistency conditions, {\it is defined with respect to a choice of the corresponding nonstandard central potential}.
We may fix the standard one by setting $Z_0 |_{T^0 M} \equiv 0$ , but  similar choices for the nonstandard ones are not necessarily the best.
{\it In $\S$ 6 , Proposition 6.1' ,  we describe a consistent choice for all $Z_i$'s which is suitable for the mixed composition of central generating functions }. Such  
compositions, as we shall see, exhibit very neatly the full symplectic and affine geometry of $M$ itself,
a fact having some interesting bearings on the problems of
quantization and semiclassical analysis.

\vspace{0.3cm}
\noindent
{\it Examples:} Standard cases only. We show the explicit  
map $z[f_{\alpha}]\equiv F_{\alpha}:M\rightarrow (TM)_0$,
given by each  generating function $f_{\alpha}$, and the
canonical relation $(m_-,m_+) \in DM$. 
To get rid of factors of 2, we often rescale  and identify
$f_{\alpha}\equiv 2f$. On $I\!\!R^2$, using (3.4) and (4.3), $F$ is written as
\[ 
\hspace{3.4cm} v_p = \partial f/\partial q \ , \ v_q = -\partial f/\partial p \ ,
\ \mbox{or} \  \  
\vec{\xi}_{\alpha} = 2\vec{v} = - J\cdot [\partial
f_{\alpha}/\partial \vec{x}] \ . \hspace{2.8cm} (4.4)
\]
where $J$ is the symplectic matrix on $I\!\!R^2$. Composing with  (3.3), we have the canonical relation as : 
\[
\hspace{4.8cm} \vec{x}_{\pm} = \vec{x} \mp J \cdot [\partial f/\partial \vec{x}] =
\vec{x} \mp \frac{1}{2} J \cdot [\partial f_{\alpha}/\partial
\vec{x}] \hspace{4.3cm} (4.5)
\]  
See \cite{26}\cite{28}. On the torus, 
(4.4)-(4.5) are valid, but we  impose a {\it standard 
consistency condition} which ammounts to constraining the map (4.4)
onto $(T\mathcal{ T}^2)_0$ only : 
\[
\hspace{5cm} |\partial f/\partial q| \ , \  |\partial f/\partial p| \ < \ \pi /2 \
, \ \mbox{on} \ U\subset \mathcal{ T}^2 \ . \hspace{4.2cm} (4.6) 
\]
On $S^2$, by (3.6) and (4.3),  
$F:S^2\rightarrow (TS^2)_0$ is written in polar coordinates as :
\[
\hspace{2.5cm} v = S^{-1} (S^0(f))  \quad , \quad 
\beta = -T^{-1}\left\{\frac{S_{\theta}\partial
f/\partial{\theta}}{\partial f/\partial \varphi}\right\} + (1-Sign
(\partial f/\partial \varphi )) \pi /2 \hspace{1.7cm} (4.7)
\]
where, using the contravariant metric on
$S^2$, we define the symbols 
\[
S^0 (f) := \|df\| \equiv \sqrt{(\partial
f/\partial\theta )^2 +
((1/S_{\theta})(\partial
f/ \partial\varphi ))^2} \ , \quad
C^0 (f) := \sqrt{1-(S^0(f))^2} \ .   
\]
Notice that (4.7)  has real solutions only if $f$
satisfies the {\it consistency condition} 
\[
\hspace{5.4cm} S^0(f)\equiv \|df\|  <
1 \ \ ,  \ \  \mbox{on $U\subset S^2$} \ . \hspace{5cm} (4.8)
\]
Composing with the symmetric exponential map (3.5), we obtain      
\[
\hspace{3.3cm} \left.
\begin{array}{l} 
\theta_{\pm} = C^{-1} (C_{\theta}C^0 (f) \mp \partial f/\partial\varphi )  \\
\varphi_{\pm} = \varphi \mp T^{-1} \{S_{\theta}(\partial
f/\partial\theta )/\alpha_{\pm}\} + (1-Sign (\alpha_{\pm}))\pi /2 \ , 
\end{array}  
\right\} \hspace{2.8cm} (4.9) 
\]
\noindent
where $\alpha_{\pm} = S^2_{\theta}C^0 (f) \pm
C_{\theta} (\partial f /\partial\varphi )$ ,    
as the local expression for the  canonical 
relation $(m_-,m_+)$ which is generated by $f_{\alpha} \equiv 2f \in \mathcal{ C}^k_{I\!\!R} (S^2)$ ,
satisfying condition (4.8).
Similarly on $H^2$, by (3.6), (3.7) and the central
equation, the map $F:H^2\rightarrow TH^2$ 
is given in local polar coordinates as :
\[
\hspace{2.1cm} \mu = \tilde{S}^{-1} (\tilde{S}^0(f))  \quad , \quad 
\beta = -T^{-1}\left\{\frac{\tilde{S}_{\rho}(\partial f/\partial
\rho )}{\partial f /\partial \varphi}\right\} + (1-Sign (\partial f
/\partial\varphi )) \pi /2  \hspace{1.6cm} (4.10)
\]
which now requires no consistency condition. 
Once more,  we have defined 
\[
\tilde{S}^0(f) := \|df\| \equiv 
\sqrt{(\partial
f/\partial\rho )^2 + ((1/\tilde{S}_{\rho})(\partial f/\partial\varphi ))^2}
\ , \
\tilde{C}^0 (f) :=\sqrt{1+(\tilde{S}^0(f))^2} \ . 
\]
Finally, composing with the symmetric exponential map we have the local expression
\[
\hspace{3cm} \left.
\begin{array}{l}
\rho_{\pm} = \tilde{C}^{-1} (\tilde{C}_{\rho}\tilde{C}^0(f) \pm
\partial f/\partial\varphi )  \\  
\varphi_{\pm} = \varphi\mp T^{-1} \left\{\tilde{S}_{\rho}
(\partial f /\partial\rho ) /\gamma_{\pm}\right\} + (1-Sign
(\gamma_{\pm})) \pi /2 \ ,  
\end{array}  
\right\} \hspace{2.9cm} (4.11)
\]
\noindent
where $\gamma_{\pm} = \tilde{S}^2_{\rho}
\tilde{C}^0 (f) \pm \tilde{C}_{\rho} (\partial f/\partial\varphi)$ ,  
for the  corresponding canonical relation on  $H^2$.

\section{Central Actions and Relations}
\setcounter{equation}{0}

We saw in the last paragraph that a function $f_{\alpha}^i\in
\mathcal{ C}^k_{I\!\!R}(M)$  can locally be taken  as central
generating function
of a canonical relation $\Lambda_{\alpha}^i\subset (\widetilde{TM})_i$, 
provided (4.2') holds. This is a 
``graphical'' condition over $U\subset M\simeq \widetilde{T^iM}\subset (\widetilde{TM})_i$ 
and is therefore written with
respect to $\Lambda_{\alpha}^i$. But generically, it is
precisely this submanifold that needs to be found given $f_{\alpha}^i$.
Furthermore, it is important to distinguish  which of the various lagrangian
submanifolds in $(\widetilde{TM})_i$ do correspond to canonical
transformations on $M$, i.e. which are pre-images, under the
symmetric exponential map, of graphs over $M_-\subset DM$. Again, since
we usually start with the central generating function, from a
practical point of view we need such a  distinction, as well as an
alternative to condition (4.2'), written directly in terms of
$f_{\alpha}^i$. We now proceed in this direction, in the {\it standard} 
case. First we obtain : 

\vspace{0.3cm}
\noindent
{\bf Lemma 5.1} {\it A function $f\equiv \frac{1}{2}f_{\alpha} \in
 \mathcal{ C}^k_{I\!\!R} (M)$, 
$k\geq 2$, can locally be the standard central generating function of a
canonical relation,  via 
central equation, only if it
satisfies all consistency conditions required for the
definition of 
the map $F:M\supset U\rightarrow
(TU)_0$, $\vec{F}:m\mapsto 
\vec{v}\in (T_mM)_0$, and}
\[
\hspace{5cm} |det[\partial F^i/\partial m^j]|<\infty  \quad , \quad \forall m \in U
\subset M \ ,  \hspace{4.2cm} (5.1)
\]
{\it for any choice of local coordinates $\{m^i\}$ on $U$, $\{v^i\}$ on
$(T_mM)_0$, with $\{F^i(m)=v^i(\tau )\}$}. 

\vspace{0.2cm} 
\noindent
{\it Proof:}
Let $\tau_{\alpha}\in \Lambda_{\alpha}
\subset (TM)_0$, $P_0(\tau_{\alpha})=m_{\alpha}\in
M$, and take local coordinates 
$\{x^1,\cdots ,x^{2n}\}$ on a
neighborhood $X\subset \Lambda_{\alpha}$ of the point $\tau_{{}_{\alpha}}$, 
$\{m^1,\cdots ,m^{2n}\}$ 
on a neighborhood $U\subset M$ of the point $m_{\alpha}$ and
$\{v^1,\cdots ,v^{2n}\}$ on a neighborhood
$V\subset (T_{m_{\alpha}} M)_0$ of the point $\vec{v}_{{}_{\alpha}} = P_v (\tau_{\alpha})$, 
$P_v : N\rightarrow V$ , $N = (U\times V) \subset (TM)_0$ . 
Then, any point $\tau \in X\subset N$ is locally 
written as $\tau\equiv \{x^i (\tau )\}\equiv \{x^i\}$, $\tau \equiv
\{m^i (P_0(\tau ))$, $v^i (P_v(\tau ))\}$, or $\tau \equiv
\{m^i(P_0(\{x^j\}))$, $v^i (P_v(\{x^j\}))\}$. Now, the graphical
condition (4.2) is simply 
$[dm^i] = A_0\cdot [dx^j] \ , \ [A_0^{ij}] = [\partial m^i /\partial
x^j] \ , \
\mbox{satisfying}\quad   0 < |det (A_0)|<\infty \ , \ \mbox{on} \ \tau$,
that is $[dx^i] = A_0^{-1}\cdot [dm^j] \ , 
\ 0<|det (A_0^{-1})|<\infty \ , $
and the failure of (4.2) is written as
$|det (A_0)|= 0 \Longleftrightarrow |det (A_0^{-1})| = \infty \ .$
But since no graphical condition over $V$ is
assumed, we have only 
$[dv^i] = A_v\cdot [dx^i] \ , \ [A_v^{ij}] = [dv^i/\partial x^j] \ , \
\mbox{satisfying}\quad  |det(A_v)|<\infty \ , \ \mbox{on} \ \tau$,
and the same cannot be said of $A^{-1}_v$ because $|det(A_v)|=0$
is a real possibility. Thus, 
$|det (A_v\cdot A^{-1}_0)|\equiv |det[\partial v^i/\partial m^j]| < \infty$
is a necessary condition for  the
existence of central generating functions. Now, if such exist, then
they satisfy (4.3) and we can write the map $F$ generated by $f\in
\mathcal{ C}^k_{I\!\!R}(U)$ as $\{v^i(\tau )= F^i (m)\}$, provided $f$ 
satisfies any required central consistency condition. \ $\square$    

\vspace{0.3cm}
\noindent
{\it Examples:} 
On $I\!\!R^2$, (5.1) becomes  
$|det [\partial^2f]|\equiv |det[\partial^2f/\partial
x^i\partial x^j]|<\infty$ . On $S^2$  we get the condition
$\left| det [\partial^2f] + (C_{\theta}/S_{\theta})(\partial f /\partial\varphi )(\partial^2f/\partial\theta\partial\varphi ) 
\right| < \infty$  , on $U\subset S^2$ , where 
$det [\partial^2f] = (\partial^2f/\partial\theta^2)(\partial^2f/\partial\varphi^2)-
(\partial^2f/\partial\theta\partial\varphi )^2$ , provided  $(\theta =0,\pi )\notin
U$, and $f$ satisfies $0<\|df\|<1$.
The restriction  on $\theta$ is easily
removable by choosing a new origin for the polar coordinates on
$S^2$,  but a new local analysis is needed when  $\|df\|(m_0)=0$. Since
$\|df\|\rightarrow 0$ implies $|\vec{v}|\rightarrow 0$, we know these
points correpond to fixed points of any canonical transformation
which can be generated by $f$. Thus, a  (local) alternative
consists of expanding $f$ around $m_0$ in linear coordinates
and applying the previous flat equation, on a small neighborhood of $m_0$.
Similarly, we can use that $S_{v}\simeq v$ to get  
$\left| det [\partial^2 f] + (C_{\theta}/S_{\theta})\left\{ (\partial f /\partial\varphi )(\partial^2f/\partial\theta\partial\varphi ) -
(\partial f /\partial\theta )(\partial^2f/\partial\varphi^2)\right\} \right| < \infty \ ,$
as the local form of (5.1)  on a
smaller neighborhood $U'$ of a point $m_0 \in U$ for which $\|df\|(m_0)=0$ .
Similarly, on $U \subset H^2$, $(\rho =0)\notin U$,
we write (5.1) locally as 
$| det [\partial^2f] + (\tilde{C}_{\rho}/\tilde{S}_{\rho})(\partial f /\partial\varphi )(\partial^2f/\partial\rho\partial\varphi ) | \cdot ( \tilde{S}_{\rho} \tilde{S}^0 (f)
\tilde{C}^0(f) )^{-1} < \infty$, if $\|df\| > 0$, and 
$| det [\partial^2f] +
(\tilde{C}_{\rho}/\tilde{S}_{\rho})\left\{ (\partial f /\partial\varphi )(\partial^2 f /\partial\rho\partial\varphi )
- (\partial f /\partial\rho )(\partial^2 f /\partial\varphi^2) \right\} | \cdot (\tilde{S}_{\rho})^{-2} < \infty \ $ 
on a smaller neighborhood of the point $m_0$ for which 
$\|df\|(m_0)=0$. Or we can expand $f$ around $m_0$ in 
linear coordinates and use the flat equation.  

\vspace{0.3cm}
Lemma 5.1 provides a necessary, but not sufficient condition for the 
existence of central generating
functions, generically.  In other words,
when $\Lambda \subset (TM)_0$ is a generic canonical relation, it is
possible that $|det (A_0)|=|det (A_v)|=0$, on $\tau \in X\subset
\Lambda$. That is, we don't necessarily have a graph over either $M$ or $T_mM$, on
$\tau = (m,\vec{v})\in \Lambda$, generically. Even when restricting to canonical
transformations on $M$,  it is possible that $det (A_0)=det (A_v)=0$, 
if $dim (M)\geq 4$. To see
this, consider the simple example:~~ $M=I\!\!R^2 \times I\!\!R^2$,
$\omega = \omega_{(1)}\oplus\omega_{(2)}$, $\alpha = \mathcal{ R}^{(1)}_0\otimes id^{(2)}$.
Furthemore, (5.1) does not distinguish between canonical relations 
and transformations.  In this respect, we extend Lemma 5.1 to   

\vspace{0.3cm}
\noindent
{\bf Lemma 5.2} {\it A
function $f\equiv \frac{1}{2}f_{\alpha}\in \mathcal{ C}^k_{I\!\!R} (M)$,  $k\geq 2$, can 
locally be
the standard central generating function of a canonical transformation
on $M$ only if it satisfies condition (5.1),  besides any
central consistency condition required for the definition of the map
$F:M\supset U\rightarrow (TU)_0$, $m\mapsto \vec{F}(m)\in (T_mM)_0$,  via the
central equation, and}  
\[
\hspace{3.7cm} 0 <|det[\partial (Exp_m (-\vec{F}(m))^i/\partial m^j]|<\infty \ \ , \ 
\ \forall m \in U\subset M \ , \hspace{2.7cm} (5.2)
\] 
{\it for any choice of local coordinates on $M$ }.

\vspace{0.2cm}  
\noindent
{\it Proof:} 
Consider $\Lambda_{\alpha} = \Phi_0^{-1}(\mathcal{ L}_{\alpha})$, where $\mathcal{ L}_{\alpha}$
is a lagrangian graph over $M_-\!\subset\!DM$. Let $\gamma = \Phi_0 (\tau
)\in Y\subset {L}_{\alpha}$, $Y$ a neighborhood of $\gamma_{\alpha}
 = \Phi_0(\tau_{\alpha})$ and consider local
coordinates $\{y^1,\cdots y^{2n}\}$ on $Y$. Similarly, take
$\{m^1_-,\cdots m^{2n}_-\}$ as local coordinates on a
neighborhood $U_-\subset M_-$ of the point
$m^{\alpha}_-=P_-(\tau_{\alpha})=P_-(\gamma_{\alpha})$. Then 
$\mathcal{ L}_{\alpha}$ is
locally a graph over $M_-\ \mbox{iff} \ [dm^i_-]=B_-[dy^i]$, where
$[B_-^{ij}]=[\partial m^i_-/dy^j]\quad \mbox{satisfies} \quad
0<|det (B_-)|<\infty \ .$
Otherwise,  if $\mathcal{ L}_{\alpha}$ is not a graph over $M_-$,
then from the definition of $B_-$~~, $det (B_-)=0$. But since 
$\Phi_0$ is a diffeomorphism, we can
rewrite above condition as
$0<|det (A_-)|<\infty, \ \mbox{where} \ [A_-^{ij}]=[\partial
m^i_-/\partial x^i] \ , $
since $A_-=B_-\cdot d{\Phi}_0$. On the other hand, if $\Lambda_{\alpha}$
is a graph over $M\simeq T^0M$, this can be further rewritten as 
$0< |det (A_-^0)| < \infty \ , \ \mbox{where}\ [(A^0_-)^{ij}] =
[\partial m^i_-/\partial m^j] \ ,$
since $A^0_-=A_-\cdot A^{-1}_0$ and $|det(A^{-1}_0)|<\infty $. But
since $m_-=P_-(\tau )=Exp_m(-\vec{v})$,  we get (5.2), provided
$\Lambda_{\alpha}$ satisfies (4.2) and  
$\tau$ is given by the central map $\vec{F}(m)=\vec{v}$, generated by
$f$ via the central equation. \ $\square$

\vspace{0.3cm}
\noindent
{\it Example:}   On $I\!\!R^2$, from
(3.3) and (4.4), (5.2) becomes 
$0 < |1+det[\partial^2f]|< \infty$,
but since (5.1) must already be satisfied, we single out the new
condition  as \ 
$det[\partial^2f] \neq -1$  \  ,  $\forall \vec{x}\in U\subset
I\!\!R^2 \ .$
However, already on $S^2$ or $H^2$, the explicit form of (5.2) in
local coordinates, for generic $f$, becomes rather long and it is
much simpler to check it directly, for each specific $f$, using the
specific expressions for $m_-(m)$ obtained from (4.9) or (4.11).

\vspace{0.3cm}
Conditions similar to (5.1) and (5.2) apply to nonstandard functions, as well. 
It is important to emphasize, once again, that any function
satisfying the conditions of Lemma 5.2 does not necessarily generate
a canonical transformation on $M$, since these conditions do not
comprise a sufficient set, generically. A more complete analysis is
needed for such characterization, which lies outside the scope of
this work. Ultimately, though, we can check explicitly the
consistency of the implicit map $m_+(m_-)$ obtained from any function
$f$ on $U\subset M$, via the central equation and the symmetric
exponential map. These functions shall be singled out :

\vspace{0.3cm}
\noindent
{\bf Definition 5.1} {\it A function $f\equiv \frac{1}{2}f_{\alpha}\in 
\mathcal{ C}^k_{I\!\!R}(M)$, $k\geq 2$, which is (locally, on $U\subset M$)
the central generating function of a
canonical transformation on $M$ is henceforth referred to simply as a
 (local) {\bf central action} on $M$. The set of all such functions is  
 denoted $\mathcal{ A}_c(U)$.
Specifically, $\mathcal{ A}^0_c (U)$ for standard
actions. More generally, a function which locally 
generates a canonical relation on $M$, via the central equation,  shall be called  
a (local) {\bf central relation} on $M$, whose set is denoted $\mathcal{ R}{el}_c(U)$} .

\vspace{0.3cm}
Of course, $\mathcal{ A}_c(U_1)\subset \mathcal{ A}_c(U_2)\ \ $ if 
$\ \ U_1\supset U_2$. Thus,  $\forall U\subset M$, $\mathcal{ A}_c(U)\supset \mathcal{ A}_c(M)$
the space of central actions on $M$, clearly non-empty since every
function $f_{\alpha} =-th$, $h\in \mathcal{ C}^k_{I\!\!R}(M)$, is a central action, for
$t$ sufficiently small. 
On the other hand, since central relations must  satisfy (5.1), but not necessarily (5.2), 
$\mathcal{ A}_c(U) \subset \mathcal{ R}{el}_c(U)$. More generally, one might still 
wish to be free from constraints imposed by central caustics. For functions on $M$, 
this is generically impossible. On the other hand, inspired by some works in symplectic topology
\cite{7}\cite{8}\cite{40}, we could allow the generating functions to be defined on $M \times I\!\!R^d$, 
instead, using the extra variables to analyse the behaviour at those critical 
points. But the study of their general definitions 
and properties is not to be found here. In an independent context, 
a very important and particular case of such ``extended'' functions 
shall be seen in $\S$ 8,9, where $d=1$.
Back to functions on $M$, our main interest, in what follows, is concerned with their 
compositions. For central actions we need only worry about the presence of caustics, 
but {\it for canonical relations we must also worry about the possibility of 
their compositions being well defined} (clean products)  \cite{12}\cite{16}\cite{43}\cite{44}. 
Thus, for simplicity, we shall not pursue on this 
broader context here, focusing instead on the compositions of central actions properly, 
from now on. But we emphasize that, {\it whenever well defined, the following 
rules of composition apply for central relations as well}.       

\vspace{0.3cm}
\noindent
{\it Illustrations:}
The simplest examples of central actions are those which generate
uniform translations on flat space. Let $\alpha \equiv T_{\vec{\xi}}:
I\!\!R^2\rightarrow I\!\!R^2$, by $\vec{x}\mapsto
\vec{x}+\vec{\xi}$, which corresponds to the lagrangian plane 
$\vec{v}=\vec{\xi}/2$, a constant, in $TI\!\!R^2$. From (4.5) one has
$\vec{\xi}=-J\cdot \left(\partial f_{\alpha}/\partial \vec{x}\right)$, integrated as 
$f_{\alpha}(\vec{x})=(J\cdot \vec{\xi} \ )\cdot \vec{x}$ , modulo
constants, rewritten as a skew-product 
$f_{\alpha}(\vec{x})=\vec{\xi}\wedge \vec{x}$ . In particular (when 
$\vec{\xi}=0$) the null action (or any constant) generates the
identity, a fact valid for every $M$. The next simplest examples, still
on flat spaces, are the homogeneous quadratic functions 
$f_{\alpha}(\vec{x})=\vec{x}^T \ B \ \vec{x}$, where $B$ is a symmetric matrix,
or equivalently $f_{\alpha}(\vec{x})=\beta p^2+\beta 'q^2+2\gamma
pq$, where $\beta ,\beta ',\gamma \in I\!\!R$. When $\gamma =0$, $\beta
=\beta '=-\tan (\lambda /2),\ f_{\alpha}$ generates a rotation by
an angle $\lambda$ through the origin. Notice that $f_{\alpha}$, as
well as (5.1), diverges everywhere when $\lambda =\pm \pi$. This
is a reflection at the origin, $\mathcal{ R}_0$,
and the corresponding lagrangian submanifold in $TI\!\!R^2$ is 
$\Lambda_{\mathcal{ R}_0}\equiv T_0I\!\!R^2$ which is not a graph over
$T^0I\!\!R^2$ anywhere. When $\beta =\beta '=0$, $\gamma =-\tanh (\lambda /2)$, 
$f_{\alpha}$ generates a pure hyperbolic
transformation with stable/unstable submanifold coinciding  with the
$p/q$ axis. Now (5.1) is always satisfied, but (5.2) fails
asymptotically as $|\lambda |\rightarrow \infty$. It is not hard to
see that every quadratic central action generates an element of 
the homogeneous symplectic group on $I\!\!R^2$ and,
conversely, every such element which can be centrally generated, is
done by a quadratic central action (Cayley transform). This is not a
property of the more familiar generating funtions of mechanics on 
$I\!\!R^2$, e.g. $f(q_-,q_+)$, for which generic translations on $M$
are also generated by quadratic functions. Adding the former
two examples, $f_{\alpha}(\vec{x})=\vec{\xi}\wedge
\vec{x}+\vec{x}^T \ B \ \vec{x}$  generate elements  of 
the inhomogeneous symplectic group. 
On $S^2$, the simplest examples are the central actions for
rotations. Here, standard case only. Let $\alpha \equiv \alpha (p,2\gamma )$ be an
element of the group of rotations $SO(3)$ acting on $S^2$,  
whose pole (fixed point) is $p$ and whose angle of rotation
is $2\gamma $.  Taking polar coordinates for $m\equiv (\theta
,\varphi )$ and   
$p\equiv (\chi ,\varepsilon )$, 
and  
$\gamma \in [-\pi /2,\pi /2]$, the  central
action  is written as  
\[
\hspace{4.5cm} f_{\alpha}(m)=-2S^{-1}\left\{S_{\gamma}\left[C_{\chi}C_{\theta}+
S_{\chi}S_{\theta}C_{(\varphi -\varepsilon )}\right]\right\}\ . \hspace{4cm} (5.3)
\]
Although it looks complicated in
local coordinates, $f\equiv \frac{1}{2} f_{\alpha}$ has the 
simple geometrical interpretation shown in Fig.5.1(a).
If we notice that $C_{\chi}C_{\theta}+S_{\chi}S_{\theta}C_{(\varphi
-\varepsilon )}=C_y$, $y=distance(m,p)$, the map $F_{\alpha}:S^2\rightarrow (TS^2)_0$ is
given by (4.7) as
\[
\hspace{2cm} \left.
\begin{array}{l}
v = C^{-1}\left\{C_{\gamma}\big/\sqrt{1-(S_{\gamma}C_y)^2}\right\} 
\equiv C^{-1}\{C_{\gamma}/C_{f}\}  \\
\beta = T^{-1}\left\{[S_{\chi}C_{\theta}C_{(\varphi -\varepsilon
)}-C_{\chi}S_{\theta}]/S_{\chi}S_{(\varphi -\varepsilon )}\right\}+
(1-Sign ((\varphi -\varepsilon ))\pi /2 
\end{array}  
\right\} \hspace{1.9cm} (5.4)
\]
\noindent
and the implicit transformation $m_+(m_-)$ generated by $f_{\alpha}$
is given by (4.9) as
\[
\hspace{2cm} \left.
\begin{array}{l}
\theta_{\pm} = C^{-1}\left\{\left[C_{\gamma}C_{\theta}\pm 
S_{\gamma}S_{\theta}S_{\chi}S_{(\varphi -\varepsilon )}\right]\big/
\sqrt{1-(S_{\gamma}C_y)^2}\right\}  \\
\varphi_{\pm} = \varphi \pm T^{-1}\left\{
S_{\gamma}\left[C_{\chi}S_{\theta}-S_{\chi}C_{\theta}C_{(\varphi
-\varepsilon )}\right]\big/\lambda_{\pm}\right\} + 
 (1-Sign(\lambda_{\pm}))\pi /2 \ , 
\end{array}  
\right\} \hspace{1.6cm} (5.5) 
\]
\noindent
where $\lambda_{\pm} = C_{\gamma}S_{\theta}\pm S_{\gamma}C_{\theta}S_{\chi}S_{(\varphi
-\varepsilon )}$ . 
Notice that when $m\rightarrow p$ or
$\bar{p}$, $v\rightarrow 0$ and $m$ is a fixed point of the
transformation (5.5), as expected, $\forall \gamma \in [0,\ \pi /2$). 
However, when $\gamma =\pi /2$ the
transformation (5.5) is not well defined. Again, in this case $\alpha
(p,2\gamma )=\alpha (p,\pi )=\mathcal{ R}_p:S^2\rightarrow S^2$, and the
corresponding lagrangian submanifold in $(TS^2)_0$ is 
$\Lambda_{\mathcal{ R}_p}\equiv (T_pS^2)_0$, which is not a graph over
$T^0S^2$ anywhere. We can also see this singularity using
(5.1), when $\gamma =\pi /2$. To see that
$m\rightarrow p$ is not a caustic singularity, for $\gamma \neq \pi
/2$, one can check directly in (5.1). Similarly, 
expanding $f$ around $m$ in linear  
coordinates $(p,q)$,  gives 
$f_{\alpha}(m)=\mbox{constant}\ +(T_{\gamma})y^2+{\it o} (y^4)\simeq (T_{\gamma})y^2$, 
with $y^2=p^2+q^2$, 
which is the central action for rotations on the plane.  

\vspace{1cm}
\begin{center}

\epsfig{file=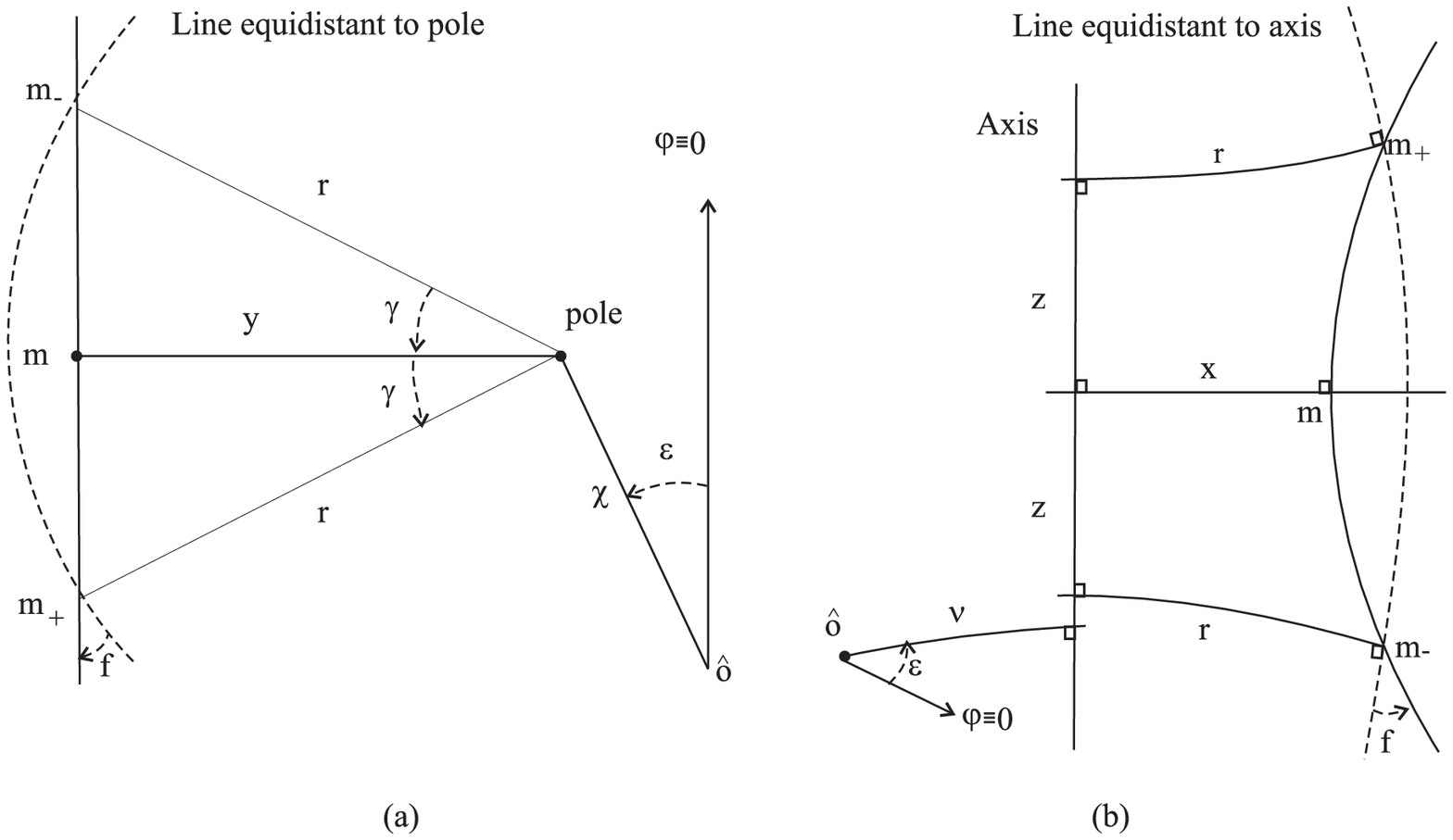,height=6.5cm}

{\footnotesize Fig. 5.1 -- Full lines represent geodesics}
\end{center}

\vspace{1cm}

\noindent
On $H^2$, we now consider the central actions for some elements
$\alpha \in SO(2,1)$. A subclass of such elements is characterized by
those   $\alpha =\alpha
(p,2\gamma )$, which are real
rotations on $H^2$ around the real pole $p\in H^2$,
through an angle $2\gamma $ , $\gamma \in \left[-\pi /2 \ , \ \pi/2\right]$. 
The corresponding central actions and
canonical transformations are analogous to the spherical ones. 
But, in opposition to the real hyperbolic rotations, stand the ideal
hyperbolic rotations. These can be characterized in two ways. We may consider $\alpha =\alpha
(\tilde{p},2\gamma)\ , \ \gamma \in \left[-\pi /2\ , \ \pi
/2\right]$, $\tilde{p}$ an ideal fixed point (a point in $I\!\!R^3$ satisfying  $x^2+y^2-z^2=1$,
while a real point in $H^2\subset I\!\!R^3$ satisfies
$z^2-(x^2+y^2)=1)$. Alternativelly, we take
 $\alpha =\tilde{\alpha}(\nu , \varepsilon ,z )$ where 
$\nu \in [0,\infty ),\ z \in (-\infty ,\infty ),\ \varepsilon
\in [0,2\pi ] ;\ (\nu ,\varepsilon )$ being the coordinates of a real
axis in $H^2$, i.e. a geodesic in $H^2$ s.t. $\nu$ is its distance to
the origin and  $\varepsilon$  is the angle this minimal geodesic
arc makes at the origin, as shown in Fig.5.1(b).
We interpret $|z |<\infty$ as a geodesic segment along this
axis, its sign determining a particular orientation for it \cite{36}. The
corresponding central action is given by 
\[
\hspace{4.8cm} f_{\tilde{\alpha}}(m)=2S^{-1}\left\{\tilde{S}_{z}\left[
\tilde{C}_{\nu}\tilde{S}_{\rho}C_{(\varphi -\varepsilon )}-
\tilde{S}_{\nu}\tilde{C}_{\rho}\right]\right\}\ , \hspace{4cm} (5.6)
\]
where $\tilde{C}_{\nu}\tilde{S}_{\rho}C_{(\varphi -\varepsilon )}-
\tilde{S}_{\nu}\tilde{C}_{\rho}=\tilde{S}_x$~~, $x=distance(m, axis)$. 
From (5.6) $f_{\tilde{\alpha}}$ exists only when
\[
\hspace{5.2cm} |\tilde{S}_{z}\tilde{S}_x|\leq 1\Leftrightarrow 0\leq x \leq
\tilde{S}^{-1} \left(1/|\tilde{S}_{z}|\right)\ , \hspace{4.7cm} (5.7)
\]
but in this neighborhood of the axis, it is a  
well defined real function whose simple geometric interpretation is shown
in Fig.5.1(b). From (4.10), the
map $F_{\tilde{\alpha}}:H^2\rightarrow TH^2$ is given by
\[
\hspace{2cm} \left.
\begin{array}{l}
\mu = \tilde{C}^{-1}\left\{\tilde{C}_{z}\big/
\sqrt{1-(\tilde{S}_{z}\tilde{S}_x)^2}\right\} 
\equiv \tilde{C}^{-1}\{\tilde{C}_z/C_f\}  \\
\beta = T^{-1}\left\{[\tilde{C}_{\nu}\tilde{C}_{\rho}C_{(\varphi -
\varepsilon )} - \tilde{S}_{\nu}\tilde{S}_{\rho}]/\tilde{C}_{\nu}S_{(\varphi -
\varepsilon )}\right\} + (1+Sign (\varphi -
\varepsilon ))\pi /2 
\end{array}  
\right\} \hspace{2.1cm} (5.8)
\]
\noindent
and the $f_{\tilde{\alpha}}$-generated intrisic transformation
$m_+(m_-)$ is given by (4.11) as
\[
\hspace{2cm} \left.
\begin{array}{l}
\rho_{\pm} = \tilde{C}^{-1}\left\{\left[\tilde{C}_{z}\tilde{C}_{\rho}\pm
\tilde{S}_{z}\tilde{S}_{\rho}\tilde{C}_{\nu}S_{(\varphi
-\varepsilon )}\right]\big/
\sqrt{1-(\tilde{S}_{z}\tilde{S}_x)^2}\right\}  \\
\varphi_{\pm} = \varphi \pm T^{-1}\left\{
\tilde{S}_{z}\left[\tilde{C}_{\nu}\tilde{C}_{\rho}C_{(\varphi
-\varepsilon
)}-\tilde{S}_{\nu}\tilde{S}_{\rho}\right]\big/\xi_{\pm}\right\} + (1+Sign (\xi_{\pm}))\pi /2 \ , 
\end{array}  
\right\} \hspace{1.7cm} (5.9) 
\] 
\noindent 
where $\xi_{\pm} = \tilde{C}_{z}\tilde{S}_{\rho}\pm \tilde{S}_{z}\tilde{C}_{\rho}
\tilde{C}_{\nu}S_{(\varphi -\varepsilon )}$ . 
The transformation (5.9) describes finite motions along lines
equidistant to the axis $(\nu ,\varepsilon )$, as shown in
Fig.5.1(b). We see that (5.9) has no real fixed point and  is also free
of central caustics, i.e. it is well defined everywhere on the same
neighborhood  of the  axis $(\nu ,\varepsilon )$ on which
$f_{\tilde{\alpha}}$ is well defined. In other words, the lagrangian
submanifold $\Lambda_{\tilde{\alpha}}\subset TH^2$ is a graph over
this  neighborhood; one can check explicitly that (5.1) does
not diverge anywhere. Notice that, as $x\rightarrow 0, \ \ m,m_+$
and $m_-$ all lie on the axis, with $z =$
distance $(m,m_+)$. However, as $|z |\rightarrow \infty$, (5.2)
fails asymptotically, in agreement with (5.7), which implies $x=0$, 
when $|z |=\infty$. 

\section{Composition of Central Actions}\setcounter{equation}{0}

Having explored the symplectic structure on central groupoids,
we now add their algebraic structure in order to answer the
following question: Let $f_{\alpha_1},f_{\alpha_2}\in
\mathcal{ C}^k_{I\!\!R} (M)$  locally be the central actions for
 two canonical transformations, respectively, 
$\alpha_1,\alpha_2:M\rightarrow M,\ m \mapsto \alpha_i(m)$. Since
$\alpha =\alpha_2 (\alpha_1)$ is another canonical transformations
on $M$, what is, locally, its central action? In other words: How do
central actions compose? We start by considering : 

\vspace{0.3cm}
\noindent
{\bf Definition 6.1} {\it Let $\zeta$ be a (local) symplectic potential for
$\omega$ on $U_{\zeta}\subset M$.  Then,
on $DU_{\zeta}\subset DM$, $\delta \zeta :=P^{\ast}_+(\zeta )-
P^{\ast}_-(\zeta )$ is a local symplectic potential  for $\delta \omega$ and 
$Z_{\zeta}:=\Phi^{\ast}_0(\delta \zeta )$ is locally a symplectic
potential for $\Omega_0$ on $W_{\zeta}\subset (TM)_0,\
\Phi_0(W_{\zeta})\subset DU_{\zeta}$, to be called an {\bf additive
potential} } .

\vspace{0.3cm}
\noindent
The reason for this name is obvious once we realize that 
\[
\hspace{5.6cm} Z_{\zeta}(\tau_1\odot \tau_2 ) \ \approx \ Z_{\zeta}(\tau_1)+Z_{\zeta}(\tau_2) \ . 
\hspace{5cm} (6.1)
\]
Here, as in Def. 3.2, this is a shorthand notation for the 
fact that the local $1$-form $Z_{\zeta} \oplus Z_{\zeta} \ominus Z_{\zeta}$
vanishes on the graph of the groupoid composition. 
Such additive potentials do not coincide
with the central potentials, i.e. $Z_{\zeta}$ does not satisfy
property (4.1), even locally, and we cannot use it to define central
generating functions directly. On the other hand, the central
potentials are not additive, as in (6.1), but in order to compose
central actions, we now realize that, on $W_{\zeta}$ , 
\[
\hspace{5cm} Z_{\zeta}=Z_0+dQ_{\zeta}\quad ,\quad Q_{\zeta}\equiv Q^0_{\zeta} \in
 \mathcal{ C}^k_{I\!\!R}(W_{\zeta}) \ . \hspace{4.1cm} (6.2)
\]
Therefore, if $\tau_i\in \Lambda_{\alpha_i}$, with
$P_0(\tau_i)=m_i\in U$, is locally generated by the standard central action 
$f_{\alpha_i}\in \mathcal{ C}^k_{I\!\!R}(M)$, then on $(TU)_0\cap W_{\zeta}$:
$Z_{\zeta}(\tau_i)=df_{\alpha_i}(m_i)+dQ_{\zeta}(\tau_i)$, combining
(4.3) and (6.2). Hence, if $\Lambda_{\alpha}$ satisfies (4.2) on
$(TU)_0$, for $\alpha =\alpha_2 (\alpha_1)$, and 
$\tau =\tau_1\odot \tau_2\in \Lambda_{\alpha}$, with $P_0(\tau )=m\in
U$, is locally generated by the standard cental action $f_{\alpha}\in
\mathcal{ C}^k_{I\!\!R}(M)$, then, on $W_{\zeta}\cap (TU)_0$ , from (6.1), 
\[
\hspace{4.5cm} df_{\alpha}(m) \ \ \approx \ \ df_{\alpha_1}(m_1)+df_{\alpha_2}(m_2)+
d\chi_{\zeta}(\tau_1, \tau_2)  \ , \hspace{3.5cm} (6.3)
\]
where we define the standard phase function $\chi_{\zeta}\equiv
\chi^0_{\zeta} :(TU)^2_0\rightarrow I\!\!R$, locally by
\[
\hspace{4.8cm} \chi_{\zeta}(\tau_1,\tau_2)=Q_{\zeta}(\tau_1)+
Q_{\zeta}(\tau_2)-Q_{\zeta}(\tau_1\odot \tau_2) \ . \hspace{3.9cm} (6.4)
\]
Now, we apply the following crucial result: 


\vspace{0.3cm}
\noindent
{\bf Proposition 6.1} {\it  
The function $\chi_{\zeta}$ defined above (6.4) 
is well defined on the whole $(TM)^2_0$ and independs on
choices of local symplectic potentials on $M$. Actually it
coincides, modulo constants, with the symplectic area of a
standard geodesic triangle on $M$ , i.e. which can be defined by
elements in $(TM)^2_0$ , determined by its midpoints $m,m_1$ and $m_2$.  
This area shall be denoted by $\Delta_0(m,m_1,m_2)$.} 

\vspace{0.2cm} 
\noindent
{\it Proof:}  
Here we rely on a mathematical construction  \cite{1}\cite{3} which consists of building an
$S^1$-principal fiber bundle over a symplectic manifold $(M,\omega )$ with
connection $\alpha$ whose curvature is $\omega  /\lambda$. We denote this as
$S^1\rightarrow (SM,\alpha )\stackrel{\pi}{\rightarrow} (M,\omega  /\lambda
)$, $d\alpha = \pi^{\ast} (\omega  /\lambda )$. Such bundle is well
defined only if $\frac{1}{2\pi\lambda} \oint_\mathcal{ B} \omega = p\in Z\!\!\!Z$,
where $\mathcal{ B}$ is any  oriented, closed two-surface without boundary
on $M$. Here, $\lambda$ is an auxiliary constant which
can be  set to zero in the end. 
To extend this construction to the double phase space [51], we
identify $SDM\equiv (\overline{SM}_-\times \overline{SM}_+)/S^1$,
$(\overline{SM}_{\pm}, \mp\alpha )$ being the same principal bundles
over $M$, but for opposite connections, and the quotient is taken with
respect to the diagonal action of $S^1\subset \mathcal{ T}^2$ on
$\overline{SM}_-\times \overline{SM}_+$. Actually, we have chosen the
connection whose curvature is $-\delta \omega/\lambda$, so we should perhaps
denote this bundle by $\overline{SDM}$, but to simplify
the notation  we  keep to $SDM$. Thus,
$S^1\rightarrow (SDM,[-\delta\alpha ])\stackrel{[\pi ]}{\rightarrow}
(DM, -\delta \omega/\lambda )$, $d[-\delta\alpha ] =  [\pi ]^{\ast}(-\delta
\omega  /\lambda )$, whose elements are denoted by $[\sigma ,\sigma ']$, where
$(\sigma ,\sigma ')\in \overline{SM}_-\times \overline{SM}_+$.  Then,
choosing identity elements of the form $[\sigma ,\sigma ]$, we extend [51]
the groupoid composition from $DM$ to $SDM$ by $[\sigma ,\sigma
']\odot [\sigma ',\sigma ''] = [\sigma ,\sigma '']$. Now, we pull
back $SDM$ using the (restricted) symmetric exponential map to get a
(trivial)  bundle over $(TM)_0$, denoted 
$(STM)_0\equiv\Phi_0^{\ast}(SDM)\stackrel{\pi_0 \times \rho_0}
{\longrightarrow} (TM)_0 \times SDM$. 
Next, consider sections $(TM)_0\rightarrow (STM)_0$  which are
obtained via the parallel transport
along the $P_0$-fibers of identity elements in $(STM)_0$, the latter
being fixed by the above choice on $SDM$. These polarized sections
$\varepsilon_0: (TM)_0\rightarrow (STM)_0$ are such that $\rho_0(\varepsilon_0(\tau)) = 
[\sigma ', \sigma '']$, where $(\sigma ', \sigma '')$ are the endpoints of a horizontal 
lift in $SM$ of the short geodesic from $m'$ to $m''$, centered on $m = P_0(\tau)$, 
for $\Phi_0(\tau) = (m', m'')$. Hence, if
$(\tau_1,\tau_2)\in (TM)_0^2$, then $\varepsilon_0 (\tau_1)\odot
\varepsilon_0 (\tau_2) = \varepsilon_0 (\tau_1\odot \tau_2) K_0
(\tau_1,\tau_2)$, where $K_0: (TM)_0^2\rightarrow S^1$ is the
holonomy in $SM$ over the triangle of composition on $M$. 
In other words, its symplectic area $A_0(\tau_1,\tau_2)$, where  
the standard triangular area function $A$ is a 
well defined continuous function on $(TM)_0^2$. 
This means that $K_0(\tau_1,\tau_2) = Exp\{\frac{\sqrt{-1}}{\lambda}A_0(\tau_1,\tau_2)\}$ is well 
defined on the whole $(TM)^2_0$ , as well as being completely independent of local
expressions for the connection $[-\delta\alpha ]$ and its pull-back $\overline{\alpha}_0$ .
But the pull-back connection $\overline{\alpha}_0$ can
locally be written as $\overline{\alpha}_0 \simeq d\theta
-\frac{1}{\lambda} Z_{\zeta}$, where $\theta$ is the fiber coordinate,
for a local choice of symplectic potential $\omega\simeq d\zeta$. 
Over the $P_0$-fibers, these can locally be rewritten as
$\overline{\alpha}_0|_m\simeq (d\theta -\frac{1}{\lambda}
(Z_{\zeta}-Z_0))|_m \simeq (d\theta -\frac{1}{\lambda} dQ_{\zeta})|_m$,
thus, provided $Q_{\zeta}|_{T^0M} \equiv 0$ and for a local
representation of identity elements as $(m,\vec{0};0)$, the
trivializing sections $\varepsilon_0$ can locally be written as
$\varepsilon_0 (\tau )\simeq (\tau ;Exp \{\frac{\sqrt{-1}}{\lambda}
Q_{\zeta} (\tau )\})$ and therefore the holonomy phase $A_0(\tau_1,\tau_2)$ is
locally identified as $\chi_{\zeta}(\tau_1 , \tau_2) = Q_{\zeta}
(\tau_1)+Q_{\zeta} (\tau_2) - Q_{\zeta} (\tau_1\odot \tau_2)$. 
On the other hand, denoting a point in $(TM)_0^2$ by $(m_1 , v_1 ; m_2 , v_2 )$, with   
$\tau_1\odot\tau_2 = \tau_3=(m_3,v_3)$,  we can almost everywhere eliminate the three vectors by 
$Exp_{m_1}(-v_1)=Exp_{m_3}(-v_3) = a$, $Exp_{m_1}(v_1)=Exp_{m_2}(-v_2) = b$, $Exp_{m_2}(v_2)=Exp_{m_3}(v_3) = c$,  and  
$\Phi_0(m_1 ,v_1) = (a, \mathcal{R}_{m_2}\mathcal{R}_{m_3}(a) = b)$, 
$\Phi_0(m_2 ,v_2) = (b, \mathcal{R}_{m_3}\mathcal{R}_{m_1}(b) = c)$, 
$\Phi_0(m_3 ,v_3) = (a, \mathcal{R}_{m_2}\mathcal{R}_{m_1}(a) = c)$, 
the exception being those midpoint triplets $\mu_{123} = (m_1,m_2,m_3)$ for which  
the fixed point set of $\mathcal{R}_{m_1}\mathcal{R}_{m_2}\mathcal{R}_{m_3}$ 
has nontrivial dimension. But for the set of such singular midpoint triplets, 
we can assume based on the lower dimensional examples, that the 
codimension of the set of singular midpoint triplets is always greater than $1$. 
In other words, the map $\Psi_0 : (TM)_0^2 \to (M\times M\times M)_{midpoints}$ 
is invertible on its immage set, except for a subset of codimension greater than $1$. 
Hence,  if $\mu_{123}$ is a regular midpoint triplet, 
$\Delta_0(\mu_{123}) = \Delta_0(m_1,m_2,m_3) = A_0(\Psi_0^{-1}(\mu_{123}))$. 
Otherwise, if $\mu_{123}$ is a singular midpoint triplet, 
consider any continuous family of regular midpoint triplets $\mu_{123}'(\epsilon)$, 
$\epsilon > 0$ s.t. $\mu_{123}'(0) = \mu_{123}$. These families always exist and, for any $\epsilon \ne 0$, 
$\Psi_0^{-1}$ is well defined and continuous, 
so that $\Delta_0(\mu_{123}'(\epsilon))$ is a continuous function of $\epsilon$. 
It follows that $\Delta_0(\mu_{123}) = \Delta_0(m_1,m_2,m_3) = \lim_{\epsilon\to 0}\{\Delta_0(\mu_{123}'(\epsilon))\}$, 
for any family  $\mu_{123}'(\epsilon) \to \mu_{123}$ .  
Therefore, the holonomy can be identified with $Exp \{\frac{\sqrt{-1}}{\lambda}\Delta_0 (m,m_1,m_2)\}$, for
$m_i=P_0 (\tau_i)$, $m=P_0(\tau_1\odot \tau_2)$, where
$\Delta_0 (m,m_1,m_2)$ stands for the symplectic area, modulo $2\pi
\lambda$, of a standard geodesic triangle with given midpoints, for which  
$\chi_{\zeta}(\tau_1 , \tau_2)$ is thus a local expression.    \ $\square$

\vspace{0.3cm}
 
Let us remark that in the simplest case of $\R^{2n}$, the midpoint
triangular area $\Delta$ is a unique well defined function on the
whole $I\!\!R^{2n}\times I\!\!R^{2n}\times I\!\!R^{2n}$. Generically, however, 
$\Delta_0$ is well defined only on a subset $U\subset M\times M\times M$, 
as exemplified below for the torus, the sphere and the hyperbolic plane. 

In every case, though, for any triplet of points  for
which $\Delta_0$ is well defined, modulo constants, we get
from propositions 4.1 and 6.1, via (6.3), the main result:  

\vspace{0.3cm}
\noindent
{\bf Theorem 6.1} {\it Let $f_{\alpha_1}, f_{\alpha_2}$ locally be standard central actions
for $\alpha_1$, $\alpha_2:M\rightarrow M$, respectively. If
$f_{\alpha}$ is locally the standard central
action for $\alpha = \alpha_2 (\alpha_1 )$, then} 
\[
\hspace{2cm} f_{\alpha}(m)\equiv f_{\alpha_1}\!\vartriangle\! f_{\alpha_2}(m)=
 Stat_{(m_1,m_2)}
\big\{f_{\alpha_1}(m_1)+f_{\alpha_2}(m_2)+\Delta_0(m,m_1,m_2)\big\} \hspace{1.6cm} (6.5)
\]
{\it defining the standard composition  of central actions on 
$M$, if $\Delta_0(m,m_1,m_2)$ is well defined, up to constants,
i.e. if $(m,m_1,m_2)$ stand as midpoints of a standard 
geodesic triangle.} 

\vspace{0.3cm}
Seen as a  product on $\mathcal{ A}_{c}^0(U)$, 
$f_{\alpha_1}\!\vartriangle
\!f_{\alpha_2}$ may not exist or if so, it may not be
unique, for $\Lambda_{\alpha}\subset (TM)_0$ may not satisfy (4.2) over $U\subset
M$, or else it
may be composed of  many sheets in $(TU)_0$. 

Furthermore, the composition rule (6.5) is not unique when
considering nonstandard central actions as well, if $M$ has closed geodesics.
In this case, repeating the steps that led to (6.3), for reducible
triples $(i,j;k)$, using (4.3') we
get the generalized  version :
\[
\hspace{4.3cm}df_{\alpha}^k (m) \approx df^i_{\alpha_1} (m_1) +
df^j_{\alpha_2} (m_2) +
d[\chi_{\zeta}]^k_{ij} (\tilde{\tau}_1,\tilde{\tau}_2) \ ,
\hspace{3.6cm} (6.3') 
\]
where $[\chi_{\zeta}]^k_{ij}$ is a local representation for
a general phase function 
$(\widetilde{TM})^2_{ij}\rightarrow I\!\!R$, given by 
\[
\hspace{4.1cm}[\chi_{\zeta}]^k_{ij} (\tilde{\tau}_1,\tilde{\tau}_2) = 
Q^i_{\zeta} (\tilde{\tau}_1) + Q^j_{\zeta} (\tilde{\tau}_2) - Q^k_{\zeta} 
(c^k_{ij}(\tilde{\tau}_1,\tilde{\tau}_2)) \ , \ \hspace{3.5cm} (6.4')
\]
with each $Q^i_{\zeta}$ being a function defined as $dQ^i_{\zeta}=
Z^i_{\zeta}-Z_i$, on appropriate subsets $W^i_{\zeta} \subset
(\widetilde{TM})_i$. As in the standard case, we identify this
function by :

\vspace{0.3cm}
\noindent
{\bf Proposition 6.1'} {\it For $(i,j;k)$ reducible, and for
appropriate choices of $Z_i , Z_j$, and $Z_k$ ,  the function
$[\chi_{\zeta}]^k_{ij}$ is well 
defined on the whole $(\widetilde{TM})^2_{ij}$, independs on local
choices of $\zeta$ and coincides, modulo constants, with the 
symplectic area of a reducible geodesic triangles on $M$
which can be defined by elements in $(\widetilde{TM})^2_{ij}\times     
(\widetilde{TM})_k$ and determined by its midpoints $(m_1,m_2;m)$, in
other words, geodesic triangles with given midpoints and sides 
which are (i,j;k)-fold geodesics. Such area shall be denoted by $\Delta^k_{ij}
(m;m_1,m_2)$ or generically by $\Delta(m,m_1,m_2)$.} 

\vspace{0.2cm}  
\noindent
{\it Proof:} 
This is an elaboration on  the proof of Prop. 6.1 . Here, we must
argue that there is a consistent choice for the $Z_j$'s, such that 
each $Q^j_{\zeta}|_{\widetilde{T^jM}}\equiv\gamma^j
:M\rightarrow I\!\!R$ is given by $\exp \{\frac{\sqrt{-1}}{\lambda}
\gamma^j (m)\} = \eta_j' (m)$,  where $\eta_j' (m)$ is the
horizontal transport over $M\simeq \widetilde{T^jM}$ of the
$L_j(m_0)$ loop holonomy, given a reference point $m_0$. Explicitly, 
if $\exp \left\{\sqrt{-1} \beta_j \right\}$ denotes the $L_j (m_0)$ holonomy in $SM$ 
(its homology class, if irreducible), then  
$\gamma^j(m)=\int_{\Sigma L'_j(m)}w$, modulo $2\pi\lambda$ and the 
constant $\beta_j$ , where   
$\partial\Sigma L'_j(m)= L'_j (m)= L_j (m_0,m)$ is the reducible loop
in $M$ given as $(m_0\rightarrow m)\circ L_j (m)
\circ (m\rightarrow m_0)\circ L^{-1}_j (m_0)$, with $L_j (m)$ denoting the $j$-fold
geodesic loop based at $m$. Now, each
$\tilde{\Phi}_j:(\widetilde{TM})_j\rightarrow DM$ provides a pull
back (trivial) bundle $(S\widetilde{TM})_j\stackrel{\rho_j}{\rightarrow}SDM$.
Taking
local trivializing sections $\tilde{\varepsilon}_j
:\widetilde{T^jM}\rightarrow (S\widetilde{TM})_j$ 
of the form $\tilde{\varepsilon}_j = \rho_i^{-1}([\sigma
,\sigma_j'])$, where $\sigma_j'=\sigma\cdot \eta_j'(m)$,  for
${\pi} (\sigma )=\pi (\sigma ')=m$, which can locally be
represented as 
$\tilde{\varepsilon}_j (\tilde{\tau}^j) \simeq (m,[\vec{v}_j];
exp\{\frac{\sqrt{-1}}{\lambda}\gamma^j (m)\})$,
then, from a similiar analysis to the standard case, 
the trivializing sections
$\varepsilon_j:(\widetilde{TM})_j\rightarrow (S\widetilde{TM})_j$
which are obtained from $\tilde{\varepsilon}_j$ via parallel
transport along the $P_j$-fibers are such that $\rho_j(\varepsilon_j(\tilde{\tau})) = 
[ \sigma ', \sigma '']$, where $(\sigma ', \sigma '')$ are the endpoints of a horizontal 
lift in $SM$ of the $j$-fold geodesic from $m'$ to $m''$, centered on $m = P_j(\tilde{\tau})$, 
for $\tilde{\Phi}_j(\tilde{\tau}) = (m', m'')$. Hence, 
if $(\tilde{\tau}_1,   
\tilde{\tau}_2)\in (\widetilde{TM})_{ij}^2$, then
$C^k_{ij} (\varepsilon_i (\tilde{\tau}_1), 
\varepsilon_j (\tilde{\tau}_2)) =   \varepsilon_k (c^k_{ij} 
(\tilde{\tau}_1,\tilde{\tau}_2))\cdot K^k_{ij}  
(\tilde{\tau}_1,\tilde{\tau}_2)$, where 
$C^k_{ij}: (S\widetilde{TM})_{ij}^2 \rightarrow (S\widetilde{TM})_k$
is the ``extension'' of the mixed composition  $c^k_{ij}$, defined as
$C^k_{ij}(z_1,z_2):= \rho^{-1}_k (\rho_i (z_1)\odot \rho_j (z_2))$, 
$z_i\in (S\widetilde{TM})_i$, etc... and $K^k_{ij}
(\tilde{\tau}_1,\tilde{\tau}_2) \in S^1$ is the holonomy in $SM$ over the
$(i,j;k)$-triangle of composition, which, if $(i,j;k)$ is reducible, 
corresponds to the symplectic area of this triangle. Since any reducible $(i,j;k)$-geodesic 
triangle stands in a bijective correspondence with a standard one, 
it follows that, as in the standard case, the $(i,j;k)$-triangular area is a well defined 
function $\Delta^k_{ij}$ of the midpoints. Therefore, the holonomy 
can be identified as $Exp\{\frac{\sqrt{-1}}{\lambda} \Delta^k_{ij}
(m;m_1,m_2)\}$, for $m_1=P_i(\tilde{\tau}_1)$,
$m_2=P_j(\tilde{\tau}_2)$ and $m=P_k(c^k_{ij}
(\tilde{\tau}_1,\tilde{\tau}_2))$, where $\Delta^k_{ij} (m;m_1,m_2)$
stands for the symplectic area, modulo $2\pi\lambda$, of the  
reducible geodesic triangle with given midpoints and sides which are  
($k$; $i$, $j$)-fold geodesics. 
This also means that $K^k_{ij}(\tilde{\tau}_1,\tilde{\tau}_2) = 
Exp\{\frac{\sqrt{-1}}{\lambda} \Delta^k_{ij}(m;m_1,m_2)\}$ 
is well defined on the whole $(\widetilde{TM})^2_{ij}$
and independs on local expressions for the connection $[-\delta\alpha
]$ and its pull-backs $\overline{\alpha}_j$. But, from the
form of such pull-back connections and for
$Q^j_{\zeta}|_{\widetilde{T^jM}}\equiv \gamma^j$ and $\varepsilon_j|_{
\widetilde{T^jM}}\equiv \tilde{\varepsilon}_j$, as above, it follows
that the trivializing sections $\varepsilon_j$ can locally be written
as $\varepsilon_j (\tilde{\tau})\simeq (\tilde{\tau};
exp\{\frac{\sqrt{-1}}{\lambda} Q^j_{\zeta} (\tilde{\tau})\})$ and therefore
$[\chi_{\zeta}]^k_{ij}$ is a local representation for $\Delta^k_{ij}
(m;m_1,m_2)$, modulo $2\pi\lambda$.   \ $\square$

\vspace{0.3cm} 
The appropriate choices of central potentials referred to in the statement are explained in its proof.  
Loosely speaking, each $Z_j$ is chosen as the pull-back of a single $1$-form $Z$ on $TM$, 
for each immersion $(\widetilde{TM})_j \rightarrow TM$ . 
Using  (6.3') we get: 

\vspace{0.3cm}
\noindent
{\bf Theorem 6.1'} {\it Let  $f^i_{\alpha_1}$,
$f^j_{\alpha_2}$ locally be central actions for
$\alpha_1,\alpha_2:M\rightarrow M$, respectively. If $(i,j;k)$ is
reducible and $f^k_{\alpha}$
is locally a central action for $\alpha =\alpha_2 (\alpha_1)$, then
it is given by} 
\[
\hspace{2cm}f^k_{\alpha}(m)\equiv f^i_{\alpha_1} \!\vartriangle^k\! f^j_{\alpha_2}  = Stat_{(m_1,m_2)} \{f^i_{\alpha_1}(m_1) + 
f^j_{\alpha_2} (m_2) + \Delta^k_{ij} (m;m_1,m_2)\} \ \hspace{1.8cm} (6.5')
\]
{\it defining the $(i,j;k)$-composition of local central actions on $M$,
provided $(m_1,m_2;m)$ can stand as midpoints for a reducible
$(i,j;k)$-geodesic triangle and each action is defined  with respect to the choices of central potentials 
referred to and explained above }.

\vspace{0.3cm}

Clearly, (6.5) is a particular case of (6.5'), with
$\Delta^0_{00}\equiv \Delta_0$. 
However, by restricting to the standard composition we have
obtained a uniquely defined rule. In extending to generic central actions this rule ceased to
be unique, even though each instance is well defined. Thus, for simplicity, when
considering multiple compositions, in $\S$ 7, we keep to the
standard case. Finally we remind that, {\it whenever meaningful (clean product \cite{44}), (6.5')  
applies for the composition of central relations on $M$}. 

\vspace{0.3cm}
\noindent
{\it Examples:} Standard cases, only. In the simplest spaces,
we provide the expressions for $\Delta$ and analyse
the specific nontrivial restrictions which apply, in each case. First : 
\[
\hspace{4.7cm} \Delta \left(\vec{x},\vec{x}_1,\vec{x}_2\right)=2 
\left\{\vec{x}\wedge \vec{x}_1+\vec{x}_1\wedge \vec{x}_2+
\vec{x}_2\wedge \vec{x}\right\} \hspace{4.1cm} (6.6)
\]
is the midpoint triangular area on $I\!\! R^2$ defining, via (6.5),  
the composition of two central
actions on  $I\!\! R^2$.    
On $\mathcal{ T}^2$, we add  
the restriction on the lengths of the triangles, 
in terms of midpoints :   
\[
\hspace{3.5cm} |q_i-q|\ , \ |p_i-p|\ , \ |q_1-q_2| \ , |p_1-p_2| <
\pi /2\quad , \quad i=1,2  \ \ . \hspace{2.8cm} (6.7)
\]
On the $2$-sphere,  we have (see Appendix ) :  
\[
\hspace{6cm} \Delta = 2 \mbox{Arg}\left\{\sigma \sqrt{1-D^2} + iD\right\} \hspace{5.2cm} (6.8)
\]
where $\sigma \equiv \sigma (m,m_1,m_2)\in \{-1,0,1\}$ has the same
sign as any of the scalar products $(\hat{m}\cdot
\hat{m}_1)$, $(\hat{m}\cdot \hat{m}_2)$, $(\hat{m}_1\cdot\hat{m}_2)$, and
$D\equiv D (m,m_1,m_2) = det [\hat{m},\hat{m}_1,\hat{m}_2]$. Here, 
$\hat{m}\in S^2 \subset I\!\!R^3$.  
For small triangles, i.e. $|\Delta |< \pi$, (6.8) simplifies to 
$\Delta (m,m_1,m_2)=2S^{-1}\left(\det \left[\hat{m},\hat{m}_1,\hat{m}_2\right]
\right)$.
This explicit form for $\Delta$ must be
placed in (6.5), provided the
triplet $(m,m_1,m_2)$ stand as midpoints of a spherical triangle with
all sides shorter than $\pi$. This nontrivial restriction : 
\[
\hspace{4.1cm} sign((\hat{m}\cdot \hat{m}_1))=sign((\hat{m}_1\cdot \hat{m}_2))=sign((\hat{m}_2\cdot \hat{m})) 
\hspace{3.5cm} (6.9)
\]
is the constraint that must be imposed on
(6.5) and (6.8) to define the standard composition of two
central actions on the sphere.  
Constraints analogous to (6.9) or (6.7) apply to each mixed
composition, for every space with closed geodesics. 
The noncompact hyperbolic plane, though similar to
the sphere, presents one subtle distinction. Here we have, for $m\simeq \vec{m}\in H^2\subset I\!\!R^3$ ,  
\[
\hspace{5cm} \Delta (m,m_1,m_2)=2S^{-1}\left(\mbox{det}\left[\vec{m},\vec{m}_1,\vec{m}_2\right]\right) \ . 
\hspace{4.1cm} (6.10)
\]
But now we notice that (6.10) only makes
sense on the subset of $H^2\times H^2\times H^2$ determined by :  
\[
\hspace{6.4cm} |\mbox{det}\left[\vec{m},\vec{m}_1,\vec{m}_2\right]|<1 \ , \hspace{5.7cm} (6.11)
\]
The compostion of two central actions on $H^2$ is given by 
(6.5) and (6.10), subject to constraint (6.11). 
Actually, (6.11) is a necessary and sufficient condition any triplet of points
on $H^2$ must satisfy in order to be the set of midpoints of a
hyperbolic triangle. As in the case of the sphere, there
is a nontrivial constraint in the composition of two central
actions, but in the hyperbolic case,  (6.11)
comes naturally from the definition of the area function
(6.10), being  intrinsic to the hyperbolic geometry. 
Its interpretation is that the three midpoints can never be too far apart, as
can be pictured by drawing a geodesic triangle on
the Poincar\'e disc. For the sphere,
on the other hand,   (6.9) is a
consequence of the restricted groupoid $(TS^2)_0$. In other words,
there do exist spherical triangles whose midpoints do not satisfy (6.9),
but they cannot be used in the definition of the standard composition of 
central actions (they can be used for mixed compositions),  
contrary to the hyperbolic case where, for every
triangle, (6.11) holds. Also note that, when
defining triangles by their triplets of vertices, instead of
midpoints, no non-trivial restriction exists (except for  sets of 
measure zero),  which is a consequence of the fact that the map 
$\{vertices\} \rightarrow \{midpoints\}$ is generally less trivial
than one would guess by looking only at the euclidean plane. 
See also Appendix and \cite{32}\cite{33}\cite{39}\cite{50}. 

\section{Multiple Compositions}
\setcounter{equation}{0}

So far we have seen how two central actions may compose into a new
one via (6.5'). Now, we want to generalize it for the composition of
an arbitrary number of standard central actions, whenever possible. 
This should be done with some care, so we first get : 

\vspace{0.3cm}
\noindent
{\bf Lemma 7.1} {\it Let $f_{\alpha_i}$  locally be standard central actions for
$\Lambda_{\alpha_i}\subset (TU)_0$. If $f_{\alpha_1}\!\vartriangle 
\!f_{\alpha_2}$ and $f_{\alpha_2}\!\vartriangle\! f_{\alpha_3}$ are unique
standard central actions on $U$, as well as $(f_{\alpha_1}\!\vartriangle\!
f_{\alpha_2})\!\vartriangle\! f_{\alpha_3}$ and
$f_{\alpha_1}\!\vartriangle \!
(f_{\alpha_2}\!\vartriangle\! f_{\alpha_3})$, then       
$(f_{\alpha_1}\!\vartriangle\! f_{\alpha_2})\!\vartriangle\! f_{\alpha_3} =
f_{\alpha_1}\!\vartriangle\! (f_{\alpha_2}\!\vartriangle\! f_{\alpha_3})$ is given by} 
\[
\hspace{2.1cm} f_{\alpha_1}\!\vartriangle\!
f_{\alpha_2}\!\vartriangle\! f_{\alpha_3}(m) = Stat_{(m_1,m_2,m_3) }
\left\{\sum_if_{\alpha_i}(m_i) 
 + P_4 (m,m_1,m_2,m_3)\right\} \ , \hspace{1.7cm} (7.1)
\] 
{\it $P_4$ being the smallest symplectic area of any
(oriented) quadrilateral  decomposable in triangles defined
by elements in $(TM)^2_0$ only and determined by the midpoints
$(m,m_1,m_2,m_3)$, up to constants.} 

\vspace{0.2cm} 
\noindent 
{\it Proof:}  
Suppose $f_{\alpha_1}\!\vartriangle\! f_{\alpha_2}$,
given by (6.5), is a unique standard central action on $U\subset M$, just as
$f_{\alpha_3}$. Applying (6.5) again:
$(f_{\alpha_1}\!\vartriangle\! f_{\alpha_2})\!\vartriangle\! f_{\alpha_3}(m)$ =
$Stat_{(m',m_3)} \{f_{\alpha_1}\!\vartriangle\! f_{\alpha_2}(m')       
+ f_{\alpha_3} (m_3) +$ $\Delta (m,m',m_3)\} 
 = Stat_{(m',m_3)}\{ Stat_{(m_1,m_2)}\left\{
f_{\alpha_1} (m_1) + f_{\alpha_2}(m_2) + 
\Delta (m',m_1,m_2)\right\}   
+ f_{\alpha_3} (m_3) + \Delta (m',m_3,m)\}$. 
If a unique solution exists, this rewrites as:
$Stat_{(m_1,m_2,m_3)} \{f_{\alpha_1}(m_1) + 
f_{\alpha_2}(m_2) + f_{\alpha_3} (m_3) +  
Stat_{(m')}\{\Delta (m',m_1,m_2)$ $ +
\Delta (m',m_3,m)\}\}$.
But, with $\{m_i\}$ constrained by the overall stationary
condition then, via the central equation (4.3), $\Delta (m',m_1,m_2)=:g_1(m')$
and $\Delta (m',m_3,m)=:g_2(m')$ provide well defined maps $\vec{G}_1$,
$\vec{G}_2:m'\mapsto \vec{v}_1,\vec{v}_2\in (T_{m'}M)_0$,  supposing that both $g_1$ and $g_2$ , as well as 
$f_{\alpha_1}\!\vartriangle\! f_{\alpha_2}$ and $(f_{\alpha_1}\!\vartriangle\!
f_{\alpha_2})\!\vartriangle\! f_{\alpha_3}$ are well defined unique central actions,
i.e.  both partial and complete unique solutions to the stationary
conditions exist. From the form of the central equation 
and the involutive character of the central potential: $Z_0(\tau
)=-Z_0(\bar{\tau})$, the $m'$ stationary condition implies
$\vec{G}_1(m')=-\vec{G}_2(m')$. Using the symmetric exponential map,  
we see that the two triangles composed form a single quadrilateral,
i.e. their sides centered at $m'$ are precisely opposite to each
other. In other words, we have that 
\[  
\hspace{3cm} Stat_{(m')} \left\{\Delta (m',m_1,m_2) +
\Delta (m',m_3,m)\right\} \equiv P_4 (m_1,m_2,m_3,m) \hspace{2.6cm} (7.2)
\]
where $P_4 (m_1,m_2,m_3,m)$ is the symplectic area of a standard 
quadrilateral whose midpoints are
$(m_1,m_2,m_3,m)$. Generically, the four midpoints do not determine
the quadrilateral uniquely. However, if all of its triangular
decompositions yield triangles which are defined by elements in
$(TM)^2_0$ only, then by Proposition 6.1 and (7.2), $P_4$ 
is the symplectic area of any such quadrilateral, up to
constants : any degeneracy in its specific geometry, which is a
continuous function of $m'$, does not alter the symplectic area, 
and any other ``$(TM)_0$-quadrilateral'', in the above sense,
will have the same symplectic area up to constants. \ $\square$  

\vspace{0.3cm}
The existence and uniqueness requirements are quite difficult to
assure beforehand, in general. Of course, if any $f_{\alpha_i}$ or
intermediary composition is not a central action, the triple
composition is void. On the other hand, if an intermediary
composition is not unique, say $f_{\alpha_1}\!\vartriangle\! f_{\alpha_2}
=\{g_1,g_2\}$, then we could proceed to $\{g_1\!\vartriangle\! f_{\alpha_3},
g_2\!\vartriangle\! f_{\alpha_3}\}$ in just the same manner, but the final
composition $(f_{\alpha_1}\!\vartriangle\! f_{\alpha_2})\!
\vartriangle\!  f_{\alpha_3}$
is not guaranteed to be associative,  in principle. However, if all
$\Lambda_{\alpha_i}, \Lambda_{\alpha_i\alpha_j,\cdots}$ are composed
of sheets whose one of them can be consistently singled out, for some
particular reason, as well as their corresponding central actions,
then we can apply Lemma 7.1 exclusively to this particular set. 
Carefully reiterating all steps to (7.1) gives :

\vspace{0.3cm}
\noindent
{\bf Corollary 7.1} {\it Let $f_{\alpha_i}$ locally be standard central actions for
$\Lambda_{\alpha_i} \subset (TU)_0$. If all intermediary, ordered
compositions $f_{\alpha_i}\!\vartriangle\!  f_{\alpha_{i+1}}$,
$f_{\alpha_{i}}\!\vartriangle\! (f_{\alpha_{i+1}}\!\vartriangle\!
f_{\alpha_{i+2}}),((f_{\alpha_{i}}\!\vartriangle\!
 f_{\alpha_{i+1}})\!\vartriangle\!
f_{\alpha_{i+2}})\! \vartriangle\! f_{\alpha_{i+3}}$,
etc... are unique standard central actions on $U$, then any $n$-string of
ordered compositions which is a unique standard central action equals any
other such $n$-string and is given by:}
\[
\hspace{2.5cm} f_{\alpha_1}\vartriangle f_{\alpha_2}\vartriangle \cdots \vartriangle
 f_{\alpha_n} (m)
= Stat_{(\{m_i\})} 
\left\{\sum_i f_{\alpha_i} (m_i) + P_{n+1} (m, \{m_i\})\right\} \hspace{1.7cm} (7.3)
\]
{\it $P_{n+1}(m,\{m_i\})$ being the smallest symplectic area of any
(oriented) $(n+1)$-polygon which can be triangulated by elements defined in
$(TM)^2_0$ only and determined by the midpoints $(m,\{m_i\})$}.

\vspace{0.3cm}
Again, if uniqueness fails, but a unique
set of central actions can be consistently singled out, then we can
use corollary 7.1 for these  particular standard central actions exclusively 
(see $\S$8). We should note that, with greater care, 
the above rules can be generalized to multiple mixed compositions of central 
actions and, even more generally, central relations whenever meaningful. 
 
Corollary 7.1 generalizes a previous result on $\R^{2n}$ \cite{28}.     
We remark that some of the above discussion regarding composition of midpoint triangles,  
in more general symmetric symplectic spaces, has been approached independently 
from the point of view of associativity for star products \cite{32}. 

\vspace{0.3cm}
\noindent
{\it Examples:} For the composition of three actions, the euclidean
plane presents an interesting feature. From (6.6) and (7.2), we get 
$P_4/2  = \mbox{Stat}_{(\vec{x}')}\left\{\vec{x}_1\wedge \vec{x}_2 + 
\vec{x}_3\wedge \vec{x} + \vec{x}'\wedge (\vec{x}_1-\vec{x}_2 + 
\vec{x}_3-\vec{x})\right\}$, which implies:
$\vec{x}_1-\vec{x}_2+\vec{x}_3-\vec{x} = 0$ .
This means that $(\vec{x},\{\vec{x}_i\})$ are the vertices
of a parallelogram with diagonals $\vec{x}_3-\vec{x}_1$,
$\vec{x}-\vec{x}_2$. But this is true for any quadrilateral in
$I\!\!R^2$, i.e. their  midpoints are vertices of a parallelogram.
Conversely, given any parallelogram on $I\!\!R^2$, there exists a
continuous family of circumscribed quadrilaterals whose midpoints are
the vertices of the given parallelogram. Such a family can be parametrized by one of the
vertices of each circumscribed quadrilateral or, equivalently, by the
midpoint of one of its diagonals, as $\vec{x}'$ above. In accordance
with (7.2), the symplectic  area independs on $\vec{x}'$, being
uniquely given as twice the area of the inscribed parallelogram:
 $P_4(\vec{x},\vec{x}_1,\vec{x}_2,\vec{x}_3)=2(\vec{x}_1\wedge \vec{x}_2 +
\vec{x}_3\wedge \vec{x} = \vec{x}\wedge \vec{x}_1 + 
\vec{x}_2\wedge \vec{x}_3)$,
which, inserted in (7.1), defines the composition of three central actions
on $I\!\!R^2$, with $(\vec{x},\{\vec{x}_i\})$ subject to the parallelogram  relation and under the
necessary existence and uniqueness conditions presented in Lemma
7.1. On the torus, the same analysis and results apply, but now
subject to the extra (standard groupoid) constraint:
$|q-q_i|, |q_i-q_j|, |p-p_i|,|p_i-p_j| < \pi /2 \ , \ i,j=1,2,3.$ 
In the spherical case, on the other hand, the above quadrilateral
ambiguity is an exception, when considering  only those
quadrilaterals defined by composing elements in $(TS^2)_0$. Then, as
with spherical triangles, a near-bijection between the sets of vertices
and  midpoints allows for a unique definition of the
quadrilateral geometry either way, i.e. each
quadrilateral is uniquely determined by its vertices or its midpoints 
(with a few exceptions) 
and the only restrictions on the latter derive from the
restricted groupoid $(TS^2)_0$. Denoting $C_{ij}=\hat{m}_i\cdot
\hat{m}_j\equiv cosine(distance(m_i,m_j))$, we can write the
midpoint area of convex quadrilaterals which are decomposable in small triangles, i.e.
$|\Delta |<\pi$, as
$ P_4 (m_1,m_2,m_3,m_4) =2 \sigma_{1234}\cdot C^{-1}
\{C_{12}C_{34}+C_{23}C_{41} - C_{13}C_{24}\} $, 
where $\sigma_{1234} = \pm 1$ is the orientation of
$(m_1,m_2,m_3,m_4)$ and the standard groupoid restrictions on the
midpoints of these simplest quadrilaterals
now become: 
$ D_{123} , D_{234} , D_{341} , D_{412} > 0 $ ,
where $D_{ijk}\equiv det [\hat{m}_i,\hat{m}_j,\hat{m}_k]$,
which, together with (7.1), define this simplest composition of three central actions on
$S^2$, under the existence and uniqueness assumptions of Lemma 7.1. 
On $H^2$, the midpoint area function for a
convex quadrilateral is given similarly by 
$ P_4 (m_1,m_2,m_3,m_4) = 2\sigma_{1234} \cdot C^{-1}
\{\tilde{C}_{12}\tilde{C}_{34} + 
\tilde{C}_{23}\tilde{C}_{41}-\tilde{C}_{13}\tilde{C}_{24}\} $ ,
where  $\tilde{C}_{ij} = cosh(distance(m_i,m_j))$. Again, the restrictions are intrinsic to
the hyperbolic geometry and can be obtained directly from the area function,
i.e. the convex set $(m_1,m_2,m_3,m_4)$ must satisfy  
$|\tilde{C}_{12}\tilde{C}_{34} + \tilde{C}_{23}\tilde{C}_{41} -
\tilde{C}_{13}\tilde{C}_{24}| < 1$ \ and each quadruplet of midpoints
satisfying this constraint determines a unique convex hyperbolic quadrilateral and
vice-versa (with a few exceptions, see below). Using Lemma 7.1, we obtain this simplest composition of three 
central actions on $H^2$. See Appendix  for a more detailed analysis on these quadrilateral geometries. 

\vspace{0.3cm}
Besides providing  explicit equations for  compositions of three
central actions, the previous discussions further illustrate some
kinds of constraints which the midpoints, or centers (the arguments of the
composing actions) are subject to. In this respect, the euclidean
plane presents the feature that, when the number of composing
actions is even, no restrictions apply, but when the number is odd,
there is a  degeneracy in the determination of the
$2\ell$-polygon from its midpoints, corresponding to a linear
functional restriction $g(\vec{x},\vec{x}_1,\cdots \vec{x}_{2\ell
-1})=0$ on the arguments of the composing central actions. For the
torus, one must further add the groupoid restrictions.
On the other
hand, in the nonflat cases studied, such a degeneracy is an exception and its
corresponding extra constraint is not present. When considering only
those polygons which can be obtained by iterated (standard) central groupoid
compositions, there is a near-bijection between the sets of
midpoints and vertices. On $H^2$, intrinsic restrictions on the
midpoints exist, though, which show explicitly in the midpoint area
function, whatever the number of composing actions. On $S^2$, only
the (standard) central groupoid restrictions apply, in every case. 

Another way to view degeneracies and
extra constraints is the following: Let $(m,\{m_i\}_n)$ be a candidate
for the set of midpoints for an $(n+1)$-polygon on $M$. Also, for each
$m_i$, let $\mathcal{ R}_{m_i}:M\rightarrow M$ be the corresponding involution
whose fixed point is $m_i$. Then, the existence of a circumscribed
$(n+1)$-polygon to the midpoints $(m,\{m_i\}_n)$ is equivalent to the
existence of a fixed point for the symplectomorphism $\mathcal{
P}_{n+1}\equiv \mathcal{ R}_{m_1}\cdot \mathcal{ R}_{m_2} \cdot \cdots \cdot
\mathcal{ R}_{m_n} \cdot \mathcal{ R}_m : M\rightarrow M$. Now, on
$I\!\!R^2$, when $n$ is odd $\mathcal{ P}_{n+1}$ is a translation, see
(2.1). Fixed points exist only when this translation is the identity,
in which case every point is fixed. For the sphere, on the other
hand, $\mathcal{ P}_{n+1} \in SO(3)$, $\forall n \in
I\!\!N$, and there is always a fixed point (actually two). However,
for $H^2$, $\mathcal{ P}_{n+1}\in SO(2,1)$ and there may or may not exist
a fixed point on $H^2$, $\forall n\geq 2$, but when there exists, it
is unique. The exception, in both cases, is when $\mathcal{
P}_{n+1}\equiv 1$ and for these sets of points the corresponding circumscribed 
polygon is not uniquely defined by the midpoints, but so is its area. 

We shall not present here an explicit characterization of multiple
compositions, for generic $n$, in every example. The reader is
referred to \cite{26}, for the euclidean case. Instead, in the
next paragraph we study a particular limit for (7.3), as $n\rightarrow  
\infty$.

\section{The Central Variational Principle}
\setcounter{equation}{0}

We now focus on the relationship between finite and infinitesimal
canonical transformations in the central description, i.e., on the
relation between finite and infinitesimal central actions. To this
end, consider 
$(\alpha )_T\equiv \{\alpha_y^{(x)}\}_T$, a continuous sequence of
canonical transformations on $M$ for which the following properties hold:
$\forall
x,y,z,t \in [0,T]$ s.t. $x+y+z =t$~, $\alpha_y^{(x)}:M\rightarrow M$
satisfies $(\alpha_y^{(x)})^{\ast}\omega =\omega$,
$\alpha_t\equiv \alpha^{(0)}_t = \alpha^{(x+y)}_z (\alpha_y^{(x)} \cdot
\alpha^{(0)}_x) = \alpha_z^{(x+y)}\cdot \alpha_{x+y}^{(0)} =
(\alpha_z^{(x+y)}\cdot \alpha^{(x)}_y)\cdot \alpha_x^{(0)} =
\alpha^{(x)}_{y+z} \cdot \alpha^{(0)}_x$ and also
$\alpha_{t+\varepsilon} = \alpha_{\varepsilon}^{(t)} \cdot \alpha_t = \alpha_t + {\it o} (\varepsilon )$, as
$\varepsilon\rightarrow 0$, with $\alpha_{\varepsilon}^{(t)}\rightarrow
id :M\rightarrow M$, where   $\alpha_{\varepsilon}^{(t)}$ is the
infinitesimal canonical transformation
defined by the
hamiltonian function $h(t)$, via Hamilton's equation or, equivalently,
by the infinitesimal central action
$f_{\alpha_{\varepsilon}^{(t)}}=-\varepsilon h(t)$, via the central equation
(4.3). We assume that $h(t)$ is a continuous function of $t$, but 
 $h(t)\neq h(t')$, in general. In other words, $h$ is a nonautonomous
hamiltonian, i.e. a continuous function on $M\times [0,T]$. 
Accordingly, we denote $h(0)\equiv h$ and, if $h(t)=h$, $\forall t
\in [0,T]$, we say that $h$ is autonomous. In this particular case,
$h(t)\equiv h$, we have 
that, for $t\in [-T,T]$,  $(\alpha_t)^{-1}\equiv \alpha_{-t}$. But generally, i.e.
$h(t)\neq h(t')$, such simplest inversibility relation only applies for very
short intervals of time, i.e.
$(\alpha_{\varepsilon}^{(t)})^{-1}=\alpha^{(t)}_{-\varepsilon}$ only in
the limit $\varepsilon\rightarrow 0$. In other words, the flow of
$\{\alpha_t\}_T$ is locally hamiltonian (autonomous), 
but not globally. 
Now, we will seek local
central actions for $\alpha_t$ given in terms of $h(t)$. To
achieve this goal we shall use the results from the previous paragraph,
but, in doing so, we should certify that $f_{\alpha_t}$ exists and
is unique, $\forall t' \in [0,t)$, in principle. Actually, both
conditions can be relaxed in this particular case, as is shown below.
Then, we obtain:

\vspace{0.3cm}
\noindent
{\bf Theorem 8.1} {\it Let $(\alpha ) _T$ be a continuous sequence of
canonical transformations on $M$, as above, where   as
$\varepsilon\rightarrow 0$, $\alpha_{\varepsilon}^{(t)}$ is the
infinitesimal canonical transformation generated by the hamiltonian
$h(t)\equiv h(m,t)$ continuous in $t$. Then, wherever the
central action for $\alpha_t\equiv \alpha_t^{(0)}$ exists,   
$f_{\alpha_t}(m) \equiv \Psi_h^t(m)\equiv \Psi_h
(m,t)$, it satisfies the {\bf Central Variational Principle:} }
\[
\hspace{3.7cm} \Psi_h (m,t) = Stat_{(\nu )} \left\{-\int_{\nu} h(m'(t'),t') dt' + 
\mathcal{ S}\!\!\!\!/_{\nu} \omega \right\}(m,t) \ , \hspace{3.1cm} (8.1)
\]
{\it for a family of continuous paths $\nu : [0,t]\rightarrow M$
geodesically centered on  $m$, where, by
definition $\{\mathcal{ S}\!\!\!\!/_{\nu} \omega\}(m,t)
\equiv [\mathcal{ S}\!\!\!\!/_{\nu} \omega]^t(m)$ is the symplectic area between
the curve $\nu$ and the geodesic from $\nu (t)$ to $\nu (0)$
centered on $m$. This area function is well
defined, up to constants, provided the geodesic is such that the full
closed circuit reducible. Furthermore, the stationary
paths $\nu$ solving (8.1) coincide with the classical trajectories on
phase space describing the continuous evolution from $\nu (0)$ to
$\nu (t)$ }.

\vspace{1cm}

\begin{center}

\epsfig{file=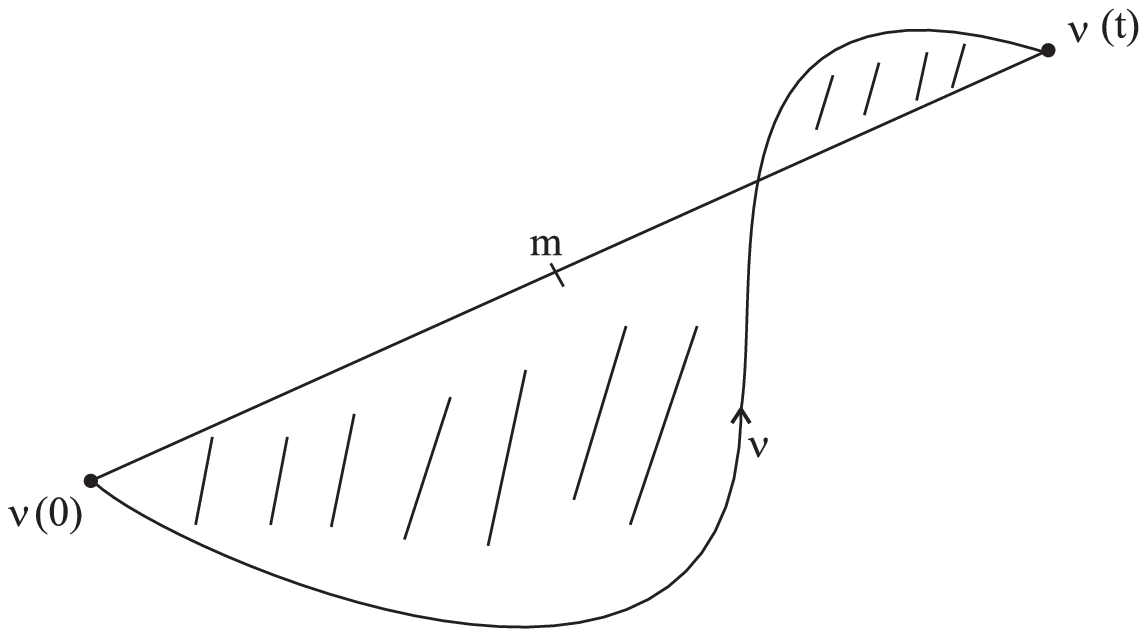,height=4.5cm} 

\vspace{0.3cm}

{\footnotesize Fig. 8.1 -- Central symplectic area of a curve}
\end{center}

\vspace{1cm} 

\noindent 
{\it Proof:} 
We start by considering those central actions which are continuously
deformed from the hamiltonian, i.e. $f_{\alpha_t}\rightarrow -
\varepsilon h$, as $t\simeq  \varepsilon \rightarrow 0$, continuously
on $U\subset M$. Thus, let $\{\Lambda_{\alpha_t}\}$ be a continuous
set of 
lagrangian submanifolds in $((TM)_0,\Omega_0)$, evolving from the
zero section $T^0M\equiv \Lambda_{\alpha_0}\simeq M$, s.t. $\forall\tau
\in\Lambda_{\alpha_t}$, $\Phi (\tau )= (m_-,m_+)$ satisfies $m_+=\alpha_t
(m_-)$. Suppose that over $\overline{U}\subset M$ there are $\ell$
branches of $\Lambda_n\equiv \Lambda_{\alpha_n}$, where
$\alpha_n\equiv \alpha_t$ for $t=t_n\in (0,T]$, generated by $\ell$
central actions $f^j_{\alpha_n}, j=1\cdots ,\ell$. Generically,
$\overline{U}$ is an open neighborhood bounded by the sets of central
caustics $\{m_n\}^j, j=1,\cdots \ell -1$.   
 Further, let
$t_k\in (0,t_n)$ be the time when a central catastrophe first appear
for $\{\Lambda_{\alpha_t}\}$, i.e. $\Lambda_{\alpha_t}$ satisfies
(4.2) everywhere on $M$, for $t<t_k$ only. Denote $\{m_k\}$ the set
of central caustics for $\Lambda_k$, then 
we can find neighborhoods in $M$ divided by $\{m_k\}$. This is
not so clear when codim$(\{m_k\})>1$, but following the evolution of
these caustics we can find appropriate subdivisions in $M$. Let's
denote by $U^1_k$ one of such neighborhoods, containing $\bar{U}$.
Then $\mathcal{ A}_c^0 (\bar{U})\supset \mathcal{ A}_c^0 (U^1_k)$. Let
$f_{\alpha_k}^1 \in \mathcal{ A}_c^0 (\bar{U})$ be the central action for
$\alpha_k$ on $\bar{U}\subset U^1_k\subset M$. Now, of all the
branches of $\Lambda_n$, over $\overline{U}$, only one is a
continuous deformation from $\Lambda_k$, over $\overline{U}$,
therefore,    of all 
the central actions for $\alpha_n$ on $\bar{U}$, only one 
 is a continuous deformation from $f^1_{\alpha_k}$ on
$\bar{U}$. And since $f^1_{\alpha_k}$ is continuously deformed from
$-\varepsilon h$, even though there are more than one central actions for
$\alpha_n$ on $\bar{U}$, there is a unique central action for
$\alpha_n$ on $\bar{U}$, denoted $f^1_{\alpha_n}$, which is continuously deformed from
$f_{\alpha_{\varepsilon}} = -\varepsilon h$, $t\simeq \varepsilon
\rightarrow 0$. This result clearly independs on the number of
subsets of $\Lambda_n$ over $\bar{U}$, or equivalently, on the number
of central actions for $\alpha_n$ on $\bar{U}$. Also, the other sets
of central catastrophes for $\Lambda_n$, $\{\tau_n\}^j, j=2,\cdots
\ell -1$, 
are not connected to $\Lambda^1_n$, so $f^1_{\alpha_n}$ is actually a
continuously deformed central action on the larger  neighborhood
$U^1_n\supset \bar{U}$, $U^1_n\subset U^1_k\subset M$ being limited
only by the set of central caustics $\{m_{t'}\}^1$, $\forall t'\in
(0,t_n)$. This also means that, for such values of $t' \ , \
f^1_{\alpha_{t'}}$ exists, besides being
continuously deformed from the hamiltonian, on $U^1_n\subset M$.
We are now ready to apply corollary 7.1 to $f^1_{\alpha_t} = 
f^1_{\alpha_{t/r}^{(r-1)}}\! \vartriangle\! f^1_{\alpha_{t/r}^{(r-2)}}\!
\cdots \vartriangle\! 
f^1_{\alpha_{t/r}^{(1)}}\!\vartriangle\! f^1_{\alpha_{t/r}^{(0)}}$,
$r$ times. That is, the continuity of $f^1_{\alpha_t}$ has
substituted for uniqueness.
From the involutive character of the central potential
$Z_0$, via the central equation we have that
$f_{(\alpha_{\varepsilon})^{-1}} = f_{\alpha_{-\varepsilon}} =-
f_{\alpha_{\varepsilon}}$, which implies that
$f_{\alpha_{-\varepsilon}} = +\varepsilon h+{\it o} (\varepsilon^3)$.
Also for $f_{\alpha_{\varepsilon}^{(t')}} =
-\varepsilon h(t') + {\it o} (\varepsilon^3)$.
 Then, letting $r\rightarrow \infty$, we can 
safely make the approximation $f^1_{\alpha_{t/r}^{(i)}}=
-{\frac{t}{r}}h(t'_i) 
+ {\it o} ((t/r)^3) \stackrel{\sim}{\rightarrow} -
 {\frac{t}{r}} h(t'_i)$,
 to get on $U^1_t\subset M$ :
\[
\hspace{2.4cm} f^1_{\alpha_t} (m) = \lim_{r\rightarrow \infty}
\left[\mbox{Stat}_{(\{m_i'\}_r)} 
\left\{\sum^r_{i=1} \left(\frac{-t}{r}\right) h (m_i',t'_i) +
P_{r+1} (m,\{m_i'\}_r)\right\}\right]  \ , \hspace{1.8cm} (8.2) 
\]
where $m'_i=m'(t'_i)$, $t'_i< t'_{i+1}\in [0,t]$. Now we realize that
(8.2) takes on the form (8.1) as we identify $\nu \equiv
\lim_{r\rightarrow \infty} (\{m'(t_i')\}_r):[0,t]\rightarrow U^1_t
\subset M$ as a continuous curve satisfying $P_0(\Phi_0^{-1} (\nu
(0), \nu (t))) = m$.  
In passing from (8.2) to (8.1), this area is integrated via a
limit $(r\rightarrow \infty)$ of the area of $(r+1)$-polygons whose
midpoints are $(m,\{m_i'\}_r)$, as $r$ of the sides tend to length
zero while the other tends to the geodesic from $m_r'$ to $m_1'$
centered on $m$.  In
order to see that the stationary path is the classical trajectory
with endpoints geodesically centered on $m$, we notice that each small
side of the $(r+1)$-polygon is a geodesic which, in the limit of very
short time intervals, i.e. of very small sides, coincides with the
local hamiltonian flow of $h(m'_i,t_i')$, which is centered on
$m'_i$, as discussed in $\S$ 4. Thus, taking the limit $r\rightarrow
\infty$ of $\{m'(t_i')\}_r$, we obtain a path that is everywhere
tangent to the locally hamiltonian flow, in other words, that
converges onto the classical trajectory.    
At first, (8.1)-(8.2) 
would apply only to those central actions that can be continuously
deformed from the hamiltonian $h$, on $U^1_t\subset M$. However, if
$f_{\alpha_t}$ is not of this type, it is always possible to
decompose it in the form 
$f_{\alpha_t} = f^1_{\alpha^{(t')}_{t-t'}}\!\vartriangle\!                     
f^1_{\alpha_{t'}^{(0)}} \ \mbox{or} \  
f^1_{\alpha_{t-t'-t''}^{(t'+t'')}}\!\vartriangle^j\! 
(f^1_{\alpha_{t''}^{(t')}}\!\vartriangle\! 
f^1_{\alpha_{t'}^{(0)}}) \ , \mbox{etc...}$
where each ``smaller'' component is continuously deformed from the
appropriate $h(t')$,
on each appropriate neighborhood. Thus, they can be written as
solutions to (8.1), with each $\nu ':[t',t-t']\rightarrow M$, etc...  
However,  the  laws for compositions of
central actions (6.5'), plus the fact that we are composing
central actions for the continuous sequence $(\alpha )_T$,  imply 
that the trajectories  $\nu ',\nu
''$  compose into a single continuous trajectory 
$\nu =\nu '' \circ \nu '$, for appropriate choices of 
$f^1_{\alpha_{t'}},f^1_{\alpha_{t''}^{(t')}}$, with the corresponding
areas $\mathcal{ S}\!\!\!\!/_{\nu '}w$ and $\mathcal{ S}\!\!\!\!/_{\nu ''}w$
summing up to $\mathcal{ S}\!\!\!\!/_{\nu}w$.
To see this, we notice that the stationary condition on $m'$ in (6.5')
implies, via central equation, that 
$[\mathcal{ S}\!\!\!\!/_{\nu}w]^{t'} (m')$ and $\Delta (m',m'',m)$
provide maps from $m'$ into reciprocally inverse elements in
$T_{m'}M$, which, via the symmetric exponential map, tells us that they
compose into a single geometric figure. Repeating the analysis with
$m''$, we have that   
$[\mathcal{ S}\!\!\!\!/_{\nu}w]^{t'} (m')$, $[\mathcal{
S}\!\!\!\!/_{\nu}w]^{t''} (m'')$ and  
$\Delta (m,m',m'')$ compose into 
$[\mathcal{ S}\!\!\!\!/_{\nu}w]^{t} (m)\equiv $
$\{\mathcal{ S}\!\!\!\!/_{\nu}w\}(m,t)$, for $t=t'+t''$, using the fact
that $\nu = \nu''\circ \nu '$, provided the composition is reducible,
i.e. provided the triangle of composition $\Delta \equiv
\Delta^k_{ij}$ is a reducible circuit, which means that the geodesic
from $\nu (t)$ to $\nu (0)$ centered on $m$ is such that   it closes
the trajectory $\nu$ into a reducible circuit.
And so on... it follows
that   $f_{\alpha_{t}}$ can also be written as a solution to (8.1),
even when it is not continuously deformed from the hamiltonian,
meaning that, for some $(m',t')\in M\times [0,t)$,
$f_{\alpha_{t'}}(m')$ does not exist, i.e. $m'$ is a central caustic
singularity for $\Lambda_{\alpha_{t'}}$, $t'< t$. So, even though generically
$f_{\alpha_{t'}}$ does not exist everywhere on $M$ , for $t'< t$,
and $f_{\alpha_{t}}$ is not everywhere deformed from the hamiltonian
function $h$ continuously, wherever $f_{\alpha_t}$ exists it can be written as a
solution to (8.1), with $\nu$ being a continuous trajectory, ~~~$\nu
:[0,t]\rightarrow M$, and $m$ being the center of the (short or long)
geodesic from $\nu (t)$ to $\nu (0)$ closing the trajectory into
a reducible circuit. \ $\square$ 

\vspace{0.3cm}
Theorem 8.1 generalizes to nonflat symmetric symplectic spaces the previous 
result on euclidean space \cite{26}. It is a real variational principle which is invariant
at least under general  transformations on $M$ 
preserving the affine connection and the symplectic form.
Besides, it does not require any local decomposition of the phase
space $M$ into  lagrangian subsets. That is, it is fully
adapted to the nontrivial geometry of $M$.
Furthermore, in opposition to the complex conterparts, {\it this real
variational principle has only real classical trajectories as
stationary solutions. The novel feature is that such
trajectories are constrained on their geodesic centers instead of the
more familiar (local) lagrangian coordinates of their endpoints}.
Therefore, in solving for the paths $\nu$ which are stationary in
(8.1), only the time $t$ and the center $m$ are held fixed.
Finally, the central action $\Psi_h(m,t)$ itself provides, via the
central equation, the finite transformation $\nu (0)\rightarrow \nu
(t)$ and, given its explicit relationship to the infinitesimal generators
$h(m,t)$, plus the fact that it is a real function on $M\times
[0,T]$, $\Psi_h$ can be seen as a {\it finite time extension of the
hamiltonian function}.    

\section{Temporal Evolution of Central Actions}
\setcounter{equation}{0}

We have just seen how the central variational principle provides, not
only for the classical trajectories obtained by the stationary
condition, but also for the central actions which generate finite
canonical transformations and can thus be seen as finite time
extensions of the hamiltonian functions. Now, we shall investigate
the temporal evolution of such central actions. This can be done in
two ways. 
First, we can examine  the temporal evolution of $\Psi_h(m,t)$,
for fixed $m$. The total variation of 
$\Psi_h(m,t)$ with respect to $t$ depends on the direction of
$\vec{\dot{m}}$. We have :
\[  
\hspace{4cm} \varepsilon \cdot \nabla_t (\Psi_h (m,t)) \equiv \varepsilon
\cdot \left\{\partial\Psi_h (m,t)/\partial t + 
\vec{\dot{m}} \rfloor d\Psi^t_h(m)\right\} \ , \hspace{3.1cm} (9.1)
\]
denoting $\delta t=\varepsilon$ . On the other  hand, by (8.1), 
\[
\hspace{3cm} \varepsilon \cdot \nabla_t\left(\Psi^t_h(m)\right)=-\varepsilon
\cdot h(\nu (t),t)+\varepsilon \cdot \left\{\vec{\dot{m}}\rfloor 
d[\mathcal{ S}\!\!\!\!/_{\nu}w]^t
(m)\right\}+{\it o} (\varepsilon^2)\ , \hspace{2.5cm} (9.2)
\]
since $\Psi^t_h(m)$ is stationary in $\nu$ and only terms in
$(\delta \nu )^2$  contribute, where ${\it o} \left((\delta \nu
)^2\right)\sim {\it o} (\varepsilon^2)$. In equation (9.2) we are
thus approximating the new path $\nu '_{t+\varepsilon}$ by the old one
$\nu_t$ , i.e. we consider only infinitesimal variations in the
endpoint $\nu (t)$ along the same
classical trajectory $\nu :(0,t+\varepsilon )\rightarrow
M$. Via the central equation, we have $d\Psi^t_h\equiv df_{\alpha_
t}:m\mapsto \tilde{\tau}_{\alpha_t}$, while
$d[\mathcal{ S}\!\!\!\!/_{\nu}w]^t:m\mapsto \tilde{\tau}_{\nu}$. But, by construction, 
$\tilde{\tau}_{\alpha_t}=\tilde{\tau}_{\nu}=\tilde{\Phi}^{-1}_i(\nu (0),\nu (t))$ and
therefore $d\Psi^t_h(m)=d[\mathcal{ S}\!\!\!\!/_{\nu}w]^t(m)$, where now
we identify $\Psi_h^t(m)\equiv \Psi^i_h (m,t)$ as a generic central action. From this and 
(9.1)-(9.2), we identify:
\[
\hspace{6.1cm} \partial \Psi_h(m,t)/\partial t=-h(m_{+},t) \hspace{5.6cm} (9.3)
\]
But since $m_{+}=Exp_m(+ \vec{F}_{\alpha_t}^i(m))$ , 
$\vec{F}_{\alpha_t}^i (m)\in  (\widetilde{T_mM})_i$
defined by $df_{\alpha_t}^i\equiv d\Psi^t_h$ via central equation, 
\[
\hspace{4.8cm} I\!\!H \left[\Psi^i_h(m,t)\right]:= h\left(Exp_m\left(+
\vec{F}_{\alpha_t}^i(m)\right),t\right)\  \hspace{4.1cm} (9.4)
\]
defines the functional $I\!\!H$ on $\mathcal{ A}_c(U)$. Then,  (9.3) can be
rewritten as:
\[
\hspace{5.3cm} \partial \Psi_h(m,t)/\partial t+I\!\!H [\Psi_h(m,t)]=0 \ , \hspace{5cm} (9.5)
\]
which is the central version of the {\it {\bf Hamilton-Jacobi equation} } .

\vspace{0.3cm}
\noindent
{\it Examples:}
On $I\!\!R^2$ (9.5) becomes
$\partial\Psi_h (\vec{x},t) / \partial t  \ + \ h (\vec{x} -
\frac{1}{2} \ J\cdot (\partial\Psi_h(\vec{x},t)/\partial \vec{x}),t) = 0$,
see \cite{23}\cite{28}. On $S^2$ and $H^2$, 
however, its explicit generic form in local coordinates is quite complicated 
and it is rather simpler to use (4.9) or
(4.11) to write $I\!\!H [\Psi^t_h(m)]$, for each specific $h$. As
simplest example, consider on $S^2$ the hamiltonian function
$h=-C_{\theta}$, generator of infinitesimal rotations around the
south/north axis, or poles. Using the convention $\Psi^t_h\equiv f_{\alpha_t}
\equiv 2f_t$, by (4.9) we rewrite the standard version of (9.5) as
$\partial f(\theta ,t)/\partial t -\frac{1}{2}\ C_{\theta}
\sqrt{1-\left(\partial f(\theta ,t)/\partial \theta \right)^2}=0 \ ,$
where we denoted $f_t(\theta ,\varphi )\equiv f(\theta ,t)$,
exploiting the $\varphi$-invariance of the action; and remembering that, by
rescaling $\Psi^t_h\equiv f_{\alpha_t}$, we must also rescale $h$,
and hence $I\!\! H$ by the same factor. Check explicitly that
the central action given by (5.3), with $\chi =0$ and $t=2\gamma$,
satisfies the previous equation. Similarly for generic examples. 

\vspace{0.3cm}
Finally, we notice that we could rewrite equation (9.1) as:
${\nabla}^g_t\left(\Psi^t_h(m)\right) := \partial
\Psi^t_h(m)/\partial t +\vec{v}_g\rfloor d\Psi^t_h(m)$, 
for
$\vec{\dot{m}}\equiv \vec{v}_g \ $ defined by
$\vec{v}_g\rfloor w =-dg$, $g\in \mathcal{ C}^k_{I\!\!\!R}(M)$. 
Using (9.5), we get:

\vspace{0.3cm}
\noindent
{\bf Proposition 9.1} {\it Let $g,h \in \mathcal{ C}^k_{I\!\!\!R}(M\times [0,T])$ and
$\Psi^t_h\in \mathcal{ A}_c(U)$, $U\subset M$, where $\Psi^t_h\equiv \Psi_{h}(t)$ is
related to $h$ via the central variational principle (8.1), for a
given time $t\in [0,T]$. The time derivative  of $\Psi^t_h$
``along $g$'', i.e. in the direction of the local hamiltonian flow of $g$,
denoted ${\nabla}^g_t (\Psi^t_h)$, is given by}
\[
\hspace{5cm} \nabla^g_t(\Psi_{h}(t)) = \left\{\Psi_h(t),g (t)\right\}-I\!\!H [\Psi_{h}(t)] \ , \hspace{4.4cm} (9.6)
\]
{\it where $\{ \ , \}$ denotes the Poisson
bracket and  the functional $I\!\!H$
is defined by (9.4) via the central equation.}

\vspace{0.3cm}
\noindent
The r.h.s. of  (9.6) involves partial derivatives in 
$M$ only.
Particularly interesting is the case where  $\Psi^t_h$ is known to
be invariant in a given direction $\vec{v}_g$ ,  
\[
\hspace{4.2cm} \nabla^g_t(\Psi_{h}(t))=0, \ \mbox{giving}
 \ \left\{\Psi_{h}(t),g(t)\right\}=I\!\!H[\Psi_{h}(t)] \hspace{3.8cm} (9.7)
\]
as a direct relationship between the functional $I\!\!H$ and the Poisson
bracket with $g$. Conversely, any function $g$
satisfying the second part of (9.7) defines curves 
$\Gamma_g:[0,T]\rightarrow U\subset
M$, along which
$\Psi^t_{h}$ is constant. The other
particularly interesting case is when 
$\left\{\Psi_h(t),g(t)\right\}=0$, for which
${\nabla}^g_t(\Psi_h (m,t))=\partial \Psi_h (m,t)/\partial t$,
computable by the Hamilton-Jacobi equation.  

\section{Conclusion}
We presented a general formalism for describing
hamiltonian systems defined on symmetric symplectic spaces, where the
local generating functions are real functions on phase space. We saw how
the central actions (relations) are defined, generating finite canonical
transformations (relations) via the central equation and the symmetric
exponential map, and how they compose via a neat formula
involving the midpoint triangular (polygonal) area. 
Then we saw how the ``extended hamiltonians'' satisfy a geometrically simple
real variational principle, which determines the classical trajectories, and 
obey a Hamilton-Jacobi equation, mixed with Poisson brackets. 
 
The authors' main motivation into this central formalism lies in its 
application to problems in quantization and semiclassical analysis,
more specifically in connection to ``Weyl quantization'' and ``star products'',
which attempt to understand the classical-quantum relationship within the 
phase space formalism. In this respect, and specially for oscillatory phenomena, it turns into a definite advantage 
the definition of real phase space generating functions,  
which can be connected to hamiltonians in such geometrical fashion, 
with their neat triangular law of composition  (see \cite{5}\cite{28}\cite{32}\cite{32.5}\cite{33}\cite{50}). 

A possible extra application refers to implementing new symplectic integrators (see \cite{25.1} for a review). 
Here, the polygonal law of composition (7.3) could be applied in the discretization process, 
making use of the fact that each 
local action for a finite small interval of time is a small, in principle 
controlled, deformation of the hamiltonian. Also, for autonomous hamiltonians the 
formalism is symmetric with respect to trajectories in both temporal directions.  

On its own, however, such geometrically simple law of composition presents new 
routes of investigation on symplectic dynamical systems.  
For,  as hamiltonian functions correspond to 
generators of infinitesimal canonical transformations, 
the actions correspond to elements in the Lie group. 
Therefore, their homogeneous presentation, which naturally extends to relations, 
sheds new light on  the canonical formalism on symmetric symplectic  spaces.  

\vspace{0.3cm}  
\noindent
{\bf Acknowledgements:} 
This is an edited english translation of the first author's PhD thesis (2000). 
PdMR thanks Jair Koiller for incentive and generous help, 
Mauricio Peixoto for an interesting remark, 
Gijs Tuynman for collaboration on various subjects related to this work, 
and Alan Weinstein for invaluable criticism, comments and suggestions. We thank CNPq for support.

\pagebreak   

\section*{Appendix} 
\addcontentsline{toc}{chapter}{Appendix}

\vspace{0.3cm}
\noindent
{\it Triangular computations} :  
In the spherical case, recall \cite{36} that 
if $\lambda_i$ are the angles of a geodesic
triangle on $S^2$, whose opposite sides are $\ell_i<\pi$,
respectively, then the following trigonometric equalities hold:
$ S_{\ell_i}/S_{\lambda_i} = constant , \ 
C_{\ell_k} = C_{\ell_i}C_{\ell_j} +S_{\ell_i}S_{\ell_j}C_{\lambda_k} \ , \ 
C_{\lambda_k} = S_{\lambda_i}S_{\lambda_j}C_{\ell_k} - C_{\lambda_i}C_{\lambda_j}$. 
Also, the area of a spherical triangle is its excess angle [36]. Now, for any spherical
triangle with sides $\ell_i<\pi$, a simple computation shows that 
$T_{\lambda_i} = K/(C_{\ell_i} - C_{\ell_j} C_{\ell_k} )$
where $\lambda_i$ is the angle opposite to $\ell_i$ and $K$ is a
constant for this triangle, $K^2\equiv
1-C_{\ell_1}^2-C_{\ell_2}^2-C_{\ell_3}^2 +
2C_{\ell_1}C_{\ell_2}C_{\ell_3}\equiv
Det^2[\hat{\alpha}_1,\hat{\alpha}_2,\hat{\alpha}_3]$, where
$\alpha_i$ are the vertices of the triangle. By correctly fixing the
orientations, we can take the $+$sign in the square root. Then, let's
denote $\ell_i=2y_i$ and $x_i\equiv distance(m_j,m_{k})$,
where $m_i$ is the midpoint of the $\ell_i$ side. We have that
$\hat{m}_i = \frac{1}{2C_{y_i}} (\hat{\alpha}_j+\hat{\alpha}_k)$, from
which,  since $C_{x_i}=\hat{m}_j\cdot \hat{m}_k$, we get that
$C_{x_i} = \frac{1}{2C_{y_k}} (C_{y_j}+C_{{z_j}}) = \frac{1}{2C_{y_i}}
(C_{y_{k}} +   C_{z_k})$,
where $z_j\equiv  distance(m_j,\alpha_j)$. From this we get:
$C_{x_1}/C_{y_1} = C_{x_2}/C_{y_2} = C_{x_3}/C_{y_3}=\Gamma, \ \mbox{a
constant}\ .$
This is a generalization of the plane trigonometric relation
$x_i/y_i=1$. Now, in order to compute $\Gamma$, we substitute the
previous equation
in the trigonometric equalities, to obtain: 
$\Gamma^2 = C^2_{x_1} + C_{x_2}^2 + C^2_{x_3} -
2C_{x_1}C_{x_2}C_{x_3} \equiv 1-Det^2[\hat{m}_1,\hat{m}_2,\hat{m}_3]$.
Using the previous equations for $T_{\lambda_i}$, $\Gamma$ and $\Gamma^2$,  we finally get 
$T_{\lambda_i} = \Gamma\sqrt{1-\Gamma^2} \big/ (\Gamma^2 -
C_{x_j}C_{x_k}/C_{x_i})\  \mbox{and}  \
T_{(\lambda_1+\lambda_2+\lambda_3)}\equiv T_{\Delta} =
\Gamma\sqrt{1-\Gamma^2}\big/(\Gamma^2-1/2)$.
Identifying $\Gamma\equiv C_{\gamma}$, we have 
$T_{\Delta}=T_{2\gamma}$, that is 
$C_{\Delta /2} =
\pm\sqrt{C^2_{x_1}+C^2_{x_2}+C^2_{x_3}-2C_{x_1}C_{x_2}C_{x_3}} \equiv
\Gamma  \ , \
S_{\Delta /2} = Det [\hat{m}_1,\hat{m}_2,\hat{m}_3]$.
The sign choice for $S_{\Delta /2}$  is fixed by the orientation.
 We still have to determine the sign of the square root in $C_{\Delta /2}$.
Obviously, if
$|\Delta |<\pi$, we must choose the $+$sign. These triangular areas
are continuously deformed from infinitesimal triangles, for which
$C_{x_i}>0$, $\forall i$. Since we are considering only short
triangles, i.e. $y_i<\pi /2$, $\forall i$, from $C_{x_i}/C_{y_i}=\Gamma$,
we get $\Gamma > 0$. On the other hand, let $|\Delta |=2\pi$, i.e. consider
$m_1,m_2,m_3$ to be collinear, same for $\alpha_1$, $\alpha_2$,
$\alpha_3$, such that the ``triangle'' coincides with a geodesic
meridian. Again, if $y_i<\pi /2$, $\forall i$, it is clear that in
this case $C_{x_i}<0$, $\forall i$, from $C_{x_i}/C_{y_i}=\Gamma$, since
$C_{\Delta /2}\equiv \Gamma <0$. And so on for triangular areas continuously deformed
from this ``big triangle''. Finally, when $|\Delta |=\pi$, $C_{\Delta
/2}=0$ and we have that, $\forall i$, $C_{x_i}=0$. In this case, $C_{y_i}$ is completely 
undetermined and so is the triangle, although the area of all such triangles is uniquely 
given by their common midpoints. It follows that
the sign of the square root is the same as the sign of the $C_{x_i}$,
or in other words $\hat{m}_j\cdot \hat{m}_k$, $\forall i,j,k$, if all
sides are short. Hence, we've got (6.8). As for the
restrictions on the midpoints, we have already obtained these. If
$C_{y_i}>0$, $\forall i$, then either $C_{x_i}>0$, $\forall i$, or $C_{x_i}=0$, $\forall i$ or else $C_{x_i}<0$,
$\forall i$. In all cases, $sign(C_{x_i}) = sign(C_{x_j})$, $\forall i,j$, which is
condition (6.9).  With some care, these analysis and results can be
extended and modified for general spherical triangles. 
Again, this previous analysis can be adapted to $H^2$,
with  care (refer to \cite{36} for hyperbolic geometry and trigonometry). 
On $H^2$ we don't have problems of antipodals or sign
choices, since $|\Delta |<\pi$, always. On the other hand, the
analogous to (6.8), namely (6.10), is well defined only when (6.11)
is satisfied. But once $\Gamma = C_{\Delta
/2}=\tilde{C}_{x_1}+\tilde{C}_{x_2}+\tilde{C}_{x_3} - 2\tilde{C}_{x_1}
\tilde{C}_{x_2}\tilde{C}_{x_3}$ is well defined, we go through the
argument backwards from $\tilde{C}_{x_1}/\tilde{C}_{y_1}=\Gamma$, to
see that the triangle is also well defined. See \cite{39} for alternative discussion.

\vspace{0.3cm}
\noindent
{\it Quadrilateral computations} :  Again we 
proceed in the spherical case and later adapt the hyperbolic
formulas. Consider a short quadrilateral with vertices
$\alpha_i$ and midpoints $m_i\equiv mid (\alpha_i,\alpha_{i+1})$, s.t.
$distance(\alpha_i,\alpha_j)<\pi$ , $\forall {i,j}\in \{1,\cdots 4\}$. This
means that not only the sides, but also the diagonals are short. Now,
denote by $m_0$ the midpoint of the diagonal $(\alpha_1,\alpha_3)$
and by $y_0$ its half length. Similarly, denote by $y_i=\frac{1}{2}
distance(\alpha_i,\alpha_{i+1})$.  Then, each of the triangles
$(\alpha_1,\alpha_2,\alpha_3)$ and $(\alpha_3,\alpha_4,\alpha_1)$ are
uniquely determined by their midpoints $(m_0,m_1,m_2)$ and
$(m_0,m_3,m_4)$, which shall be denoted by $\Delta_{12}$ and
$\Delta_{34}$, respectively, with the same notation referring to
their respective areas. Further, we denote $x_{ij}=distance
(m_i,m_j)$. From the triangular analysis, we know that
$C_{x_{12}}/C_{y_0} =C_{\Delta_{12}/2}$, $C_{x_{34}}/C_{y_0} =
C_{\Delta_{34}/2}$. Similarly for the other partition,
$C_{x_{23}}/C_{y_0'} = C_{\Delta_{23}/2}$, $C_{x_{41}}/C_{y_o'}=
C_{\Delta_{41}/2}$, where $y_0'=\frac{1}{2}$ distance
$(\alpha_2,\alpha_4)$, $m_0'$ being its midpoint, and so on. Therefore,
$C_{x_{12}}/C_{x_{34}} = C_{\Delta_{12}/2} /C_{\Delta_{34}/2} \ \ , \ \ 
C_{x_{23}}/C_{x_{41}} = C_{\Delta_{23}/2} /C_{\Delta_{41}/2}  \ . $
These equations generalize the parallelogram relation on the plane,
$x_{12}/x_{34}=1=x_{23}/x_{41}$, but contrary to the plane, 
they impose no constraint on the midpoints. To compensate for this
fact, in the spherical case the 4 midpoints uniquely determine the
area as well as the specific geometry of the short quadrilateral. In
other words $m_0=m_0 (\{m_i\})$ and similarly for $m_0'$. To see
this, let's denote  $\tau_0 = \Phi^{-1}_0 (\alpha_1,\alpha_3)$. Ie.
$\tau_0 = (m_0,\vec{v}_0)$, $|\vec{v}_0|=y_0$. Also, we denote by
$\mathcal{ R}_m$ the involution through $m$ and consider the element of
$SO(3)$ defined as $\sigma_{12}^2:=\mathcal{ R}_{m_1}\mathcal{ R}_{m_2}$.
Similarly, $\sigma_{34}^2:=\mathcal{ R}_{m_3}\mathcal{ R}_{m_4}$. Now, for any
element $\sigma \in SO(3)$, consider the vector field
${X}_{\sigma}\subset TS^2$ defined by $\tau\in {X}_{\sigma}$ iff $\Phi
(\tau ) = (m_-,m_+)$ s.t. $m_+=\sigma^2 (m_-)$. Then, 
the condition which guarantees that triangles $\Delta_{12}$ and
$\Delta_{34}$ compose into a quadrilateral $\square_{1234}$ can be
written as 
$\tau_0 \in {X}_{\sigma_{12}} \cap {X}_{\sigma_{34}^{-1}}$.
But, $\tau_0\in {X}_{\sigma_{12}}$ only if the pole $p_{12}$ of
$\sigma_{12}$ lies in the polar line of $\tau_0$, which   is defined
as the orthogonal geodesic to $\vec{v}_0$, at $m_0$. Thus, the first 
condition obtained  is
that $m_0, p_{12}$ and $p_{34}$ be collinear. That is, $Det
[\hat{m}_0,\hat{p}_{12},\hat{p}_{34}]=0$. In fact, we have more:
$\hat{p}_{12} = \frac{1}{S_{x_{12}}} \ \hat{m}_1\times \hat{m}_2$,
$\hat{p}_{34} = \frac{1}{S_{x_{34}}} \ \hat{m}_3\times \hat{m}_4$.
Therefore $Det[\hat{m}_0,\hat{m}_1\times \hat{m}_2, \hat{m}_3\times
\hat{m}_4]=0$. We can thus write
$\hat{m}_0 = \vec{z}/|\vec{z}|, \ \mbox{where} \  \vec{z} = z_{12}
(\hat{m}_1\times \hat{m}_2) + z_{34} (\hat{m}_3\times \hat{m}_4)$.
Further: denoting $\mu_{12}=$distance $(m_0,p_{12})$
and $\mu_{34} =$distance $(m_0,p_{34})$, we have $|S_{\mu_{12}}
T_{x_{12}}| = |T_{y_0}|=|S_{\mu_{34}}T_{x_{34}}|$, but since 
$|S_{\mu_{12}}S_{x_{12}}|=|\hat{m}_0\times (\hat{m}_1\times
\hat{m}_2)|$ and $|S_{\mu_{34}}S_{x_{34}}|= |\hat{m}_0\times
(\hat{m}_3\times \hat{m}_4)|$, it follows that
$|\hat{m}_0\times (\hat{m}_1\times \hat{m}_2)|\cdot|\hat{m}_3\cdot
\hat{m}_4|  = 
|\hat{m}_0\times (\hat{m}_3\times \hat{m}_4)|\cdot|\hat{m}_1\cdot
\hat{m}_2|$.
Then,  we get:
$\hat{m}_0=\vec{z}/|\vec{z}|, \ \mbox{where} \  \vec{z} = \alpha C_{x_{12}} (\hat{m}_3\times
\hat{m}_4) + \beta C_{x_{34}}  
(\hat{m}_1\times
\hat{m}_2) \ , \ \alpha,\beta\in \{-1,1\}$.
The sign choices must be made with care, but for small convex
 quadrilaterals we take the $+$ choice
twice. In this case, with $R = C_{x_{12}}C_{x_{34}} 
+ C_{x_{23}}C_{x_{41}} - C_{x_{13}}C_{x_{24}}$, we have that
$|\vec{z}|^2 \equiv z^2 = C^2_{x_{12}} +C^2_{x_{34}} -
2C_{x_{12}}C_{x_{34}} R$.  Analogous equations hold for $\hat{m}_o'$.
We have fixed the geometry of the small convex
quadrilateral uniquely from the midpoints $\{m_1,\cdots ,m_4\}$, as
mentioned earlier. The exception is when $\mathcal{R}_1\mathcal{R}_2\mathcal{R}_3\mathcal{R}_4 = 1$ , but now, 
contrary to the euclidean plane, this is really the exception, not the rule.  As for
the area, from (6.8) we have
$S_{\Delta_{12}/2} = \frac{1}{z}\left\{C_{x_{34}} - R C_{x_{12}}\right\} \ ,
\ S_{\Delta_{34}/2} = \frac{1}{z}\left\{C_{x_{12}}-RC_{x_{34}}\right\} $.
Then, in this simpler case, denoting the quadrilateral area by $P_4$, we get
$C_{P_4/2} = R$ , $P_4 = 2\sigma_{1234} \cdot C^{-1} \{ C_{x_{12}}C_{x_{34}} + C_{x_{23}}C_{x_{41}} -
C_{x_{13}}C_{x_{24}} \} \ .$
Here we have restricted to convex quadrilaterals decomposed in small triangles
(area $< \pi$). Greater care is needed for the sign
choices in the expression of $\hat{m}_0$, otherwise. As for the restrictions, in this
simpler case, by imposing $\hat{m}_0\cdot \hat{m}_i>0$, we get $D_{123} > 0$, and cyclic .
Again, in transposing to the hyperbolic plane, we don't have as many worries
about sign choices, however in this case the analogous area function,
$P_4 = 2\sigma_{1234} \cdot C^{-1}(\tilde{R})$, where $\tilde{R} = \tilde{C}_{12}\tilde{C}_{34} + \tilde{C}_{23}\tilde{C}_{41} - \tilde{C}_{13}\tilde{C}_{24}$, 
is well defined only
when $|\tilde{R}| < 1$ \  and, as in the triangular analysis,
once this holds the convex quadrilateral exists. And so on for the
general case.  
  



}

\end{document}